\documentclass  [a4paper, 10pt] {article}
\usepackage{amsmath}
\usepackage{amssymb}
\usepackage{amsfonts}
\usepackage [latin1] {inputenc}
\usepackage [T1] {fontenc}
\usepackage[french]{babel}
\newtheorem {theo}{Theorem}[section]
\newtheorem {lem} {Lemma}[section]

\makeatletter

\@addtoreset{equation}{section} \makeatother

\def\R{\mathbb{R}}

\def\N{\mathbb{N}}

\def \d {\delta}
\def \E {{\bf E}}

\def \T {\hbox{ Tr\,}}

\def \I {\hbox{Im\,}}

\def \s {\sigma}

\title{ ASYMPTOTIC PROPERTIES OF RESOLVENTS OF LARGE  DILUTE WIGNER RANDOM MATRICES }
\author{{\bf  S. Ayadi and O. Khorunzhiy}\\
LMV - Laboratoire de Math\'ematiques
de Versailles\\
 Universit\'e de Versailles - Saint-Quentin-en-Yvelines\\
78035 Versailles (FRANCE)}
\begin{document}

\maketitle

\begin{abstract} We study the spectral properties of the dilute Wigner random
real symmetric $n\times n$ matrices $H_{n,p}$ such  that
the entries $H_{n,p}(i,j)$ take zero value with probability $1-p/n$.
We prove that under rather general conditions on the probability distribution
of $H_{n,p}(i,j)$ the semicircle law is valid for
 the dilute Wigner ensemble in the limit $n,p\to\infty$.
In the second part of the paper we study the leading term of the correlation function of the resolvent $G_{n,p}(z) = (H_{n,p}-zI)^{-1}$ with large enough $\vert$Im$z\vert$ in the limit $p,n\to\infty$, $p=O(n^\alpha)$,
$3/5<\alpha<1$. We show that this leading term,
when considered in the local spectral scale, converges to the same
limit as that of the resolvent correlation function of the Wigner ensemble of random matrices. This shows that the moderate dilution of the Wigner ensemble
does not alter its universality class.

\end{abstract}

\section{Introduction}

The initial interest in the  spectral theory of large random matrices
has been motivated by the stochastic approach
to the descriptions of the energy spectrum of heavy nuclei (see
e.g. the collection of early papers \cite{Porter}). Later random matrices
of infinitely increasing dimensions
have seen numerous applications in various branches of theoretical and mathematical physics
such as
statistical mechanics of disordered spin systems, solid state
physics, quantum chaos theory, two-dimensional gravity  (see
monographs and reviews \cite{BIZ,CPV,GMW,Haake}). In mathematics, the
spectral theory of random matrices has revealed deep links with the
orthogonal polynomials, integrable systems, representation theory,
combinatorics, non-commutative probability theory and other
theories \cite{Mehta}.

The first result of the spectral theory of large random matrices
was obtained by E. Wigner in the middle of 50th \cite{W}
on the
eigenvalue distribution of the ensemble
$A_{n}$ of $n\times n$ real symmetric matrices of the form
$$
A_{n}(i,j)=\frac{1}{\sqrt{n}}a(i,j), \quad i,j=1,\dots,n,
\eqno (1.1)
$$
where $\{a(i,j), \ 1\le{i}\le{j}\le{n}\}$ are independent
 random variables.
 E. Wigner \cite{W} proved that in the case when  random variables $a(i,j)$ have symmetric probability distribution with the second moment $v^2$ and such that all  random variables $a(i,j)$ have all
moments finite,  then  the  eigenvalue counting function
$$
\sigma_{n}(\lambda,A_{n})=\frac{1}{n}
\sharp\{\lambda^{(n)}_{j}\le{\lambda}\},
\eqno (1.2)
$$
where $\lambda^{(n)}_{1}\le{\ldots}\le{\lambda^{(n)}_{n}}$ denote
the eigenvalues of $A_{n}$,
 weakly converges in average as
$n\rightarrow \infty$ to the limiting function $\sigma_{sc}(\lambda)$,
with the derivative  $\sigma_{sc}'(\lambda)=\rho_{sc}$ of the semicircle form
$$
\rho_{sc}{(\lambda)}=
\sigma_{sc}^{'}(\lambda) = {1\over 2\pi v^2}
\begin{cases} \sqrt{4v^2-\lambda^{2}}, & \textrm{if}\ \
|\lambda|\le{2v} ;
\cr
0,  &  \textrm{otherwise.}
\cr
\end{cases}
\eqno (1.3)
$$
This limiting distribution (1.4) is referred to as
the Wigner distribution and the convergence
$$
\sigma_{n}(\lambda,A_{n})\to \sigma_{sc}(\lambda)
\eqno (1.4)
$$ is known as the semicircle (or Wigner) law. Also the ensemble of random matrices
$\{A_n\}$ (1.1) with jointly independent
centered random variables $a_{ij}$ having the variance $v^2$  is called the Wigner ensemble.
Wigner has proved convergence (1.4) with the help of  the averaged  moments of
$\sigma_{n}(\lambda,A_{n})$ determined in natural way  by the traces of powers of $A_n$.

Another proof of the semicircle law (1.4)
can by obtained in frameworks of
 the resolvent approach
introduced first in the random matrix theory by \mbox{V. Marchenko} and L. Pastur \cite{MP}.
Moreover, it can be shown that the norma\-lized trace of the resolvent
 $g_{n}(z)=\frac{1}{n}\T (A-z)^{-1}$ converges to the
Stieltjes transform $w(z)$ of $\sigma_{sc}(\lambda)$ under much more
relaxed conditions than those of the Wigner's original proof \cite{MP,P}.

The further progress in the studies of the resolvent of random matrices of the Wigner ensemble is related with the asymptotic expansions of the covariance function
$$
C_{n}(z_{1},z_{2})={\bf E}\{g_{n}(z_{1})g_{n}(z_{2})\}-{\bf
E}\{g_{n}(z_{1})\}{\bf E}\{g_{n}(z_{2})\}
\eqno (1.5)
$$
that is sometimes referred to as the correlation function of the resolvent.
 In paper \cite{KKP}, it is proved that if  arbitrary distributed
random variables $a_{ij}$ have the fifth moment finite, then
 the asymptotic expansion of $C_{n}(z_{1},z_{2})$ is given by
$$
C_n(z_1,z_2)
=\frac{1}{n^{2}}f(z_{1},z_{2})+o(\frac{1}{n^{2}}), \quad \vert \I z_j\vert >2v,
\eqno (1.6)
$$
where the leading term $f(z_1,z_2)$  depends on the limiting Stieltjes
transform $w(z)$ and on the moments $\E a(i,j)^2 = (1+\delta_{ij})v^2$ and
$V_4 = {\bf E} a(i,j)^4$; the form of this term is  such that in the local scaling limit the following convergence holds
$$
\lim_{n\to\infty} {1\over n^2} f(\lambda-{r\over 2n} +i0,
\lambda+{r\over 2n}-i0) = -{1\over r^2}, \quad \vert \lambda\vert <2v.
\eqno (1.7)
$$
This expression coincides with the averaged version of the density-density covariance
function obtained by F. Dyson for the Gaussian Orthogonal Ensemble
of random matrices \cite{D}. The right-hand side of (1.7) does not depend on
the moments $V_{2l}$ of random variables and this result supports the universality conjecture for the local spectral properties of random matrices in the bulk of the spectrum.

During two last decades, there is a growing interest to
certain versions of the Wigner ensemble of random matrices named by the dilute random matrices
(for example, see the review  \cite{KKPS}).
The spectral properties of this kind of ensembles have been intensively studied
numerically and analytically in  theoretical physics literature
(see  \cite{EV,FM,RD} for the earlier results and
\cite{K,S} for the  recent advances and references).
In particular, it is shown that  in the limit of large $n$ and not too strong dilution  the semicircle law
is valid for the dilute Wigner ensembles \cite {RD}; also  the
universal behavior of the density-density correlation function
is detected on the theoretical physics level of rigour \cite{FM}.

The aim of the present paper is two-fold. First, we prove the analog
of the statement (1.5) for the dilute Wigner ensembles of random
matrices such that $a_{ij}$ belong to fairly wide classes of random
variables. To do this, we develop the cumulant expansions approach
proposed in \cite{KKP} to study the resolvent of the Wigner random
matrices. In the second part of the present paper, we use the
technique developed and prove analogs of relation (1.6) for the
dilute Wigner random matrices. We show that in certain asymptotic
regimes the analogs of the universality  relation (1.7) are true.
This allows one to conclude about the universality of the local
spectral statistics of dilute Wigner random matrices.

 The outline of this paper is as follows. In Section 2, we define the dilute
Wigner random matrix ensemble $H_{n,p}$ and formulate our main results.
 In
Section 3 we prove the semicircle law.
In Section 4 we study the corresponding
correlation function $C_{n,p}(z_{1},z_{2})$ of the resolvent
$G_{n,p}(z) = (H_{n,p}-zI)^{-1}$  we show that the variance
of the normalized trace of the resolvent ${\bf
Var}g_{n,p}(z)$ is bounded by $(np)^{-1}$; also we find the leading terms
of $C_{n,p}(z_{1},z_{2})$. In Section
5 we prove the auxiliary statements used in Section 4.
In Section 6 we study the
asymptotic properties of the leading terms of $C_{n,p}(z_1,z_2)$
and prove
analogs of relation (1.7).

 \vskip0,5cm
\section{Main results and the scheme of the proofs}

\subsection{Dilute Wigner ensemble of random matrices }

Let us consider  a family of independent Bernoulli random
variables
 ${\cal D}_{n,p}=\{d_{n,p}(i,j): \   1\le{i}\le{j}\le{n}\}$ with the law
$$
d_{n,p}(i,j) = \left\{
\begin{array}{lll}
1 & \textrm{with probability} & p/n \\
0 &  \textrm{with probability} & 1-p/n, \quad 0<p\le{n}
\end{array}\right.
$$
that is independent of the family of independent random variables
{${\cal A}_{n}$}.
We assume that
${\cal A}_{n}$ and ${\cal D}_{n}$ are defined on the same
probability space $(\Omega,{F},{\bf P})$ and we denote by
${\bf E}\{.\}$ the mathematical expectation with respect to ${\bf
P}$.

We assume that the random variables $a_{ij}$ satisfy conditions
$$
\E a_{ij} =0, \quad \E a_{ij}^2 = (1+\d_{ij})\, v^2,
\eqno (2.1)
$$
where $\d_{ij}$ is the Kronecker symbol.
In what follows, we  require the existence of several more absolute
moments of $a(i,j)$ that we denote by
$$
\mu_{r}=\sup_{1\le{i}\le{j}\le{n}}{\bf E}\{|a(i,j)|^{r}\},
$$
where the upper bound for $r$ is to  be specified.

We define the dilute Wigner ensemble as the family of   real symmetric $n\times n$
random matrices $H_{n,p}$ of the form
$$
H_{n,p}(i,j)=\frac{1}{\sqrt{p}}a(i,j)d_{n,p}(i,j), \quad
1\le{i}\le{j}\le{n}
\eqno (2.2)
$$
and consider the resolvent
$$
G_{n,p}(z)=(H_{n,p}-z)^{-1}, \quad \I z\neq{0}.
\eqno (2.3)
$$
The normalized trace of the resolvent $g_{n,p}(z) = n^{-1} \T G_{n,p}(z)$ represents the Stieltjes
transform of the normalized eigenvalue counting function
$\sigma(\lambda;H_{n,p})$ (1.2)
$$
g_{n,p}(z)=\frac{1}{n}\T G_{n,p}(z)=\int
{ d\sigma(\lambda,H_{n,p})\over \lambda-z},
 \ \I z \neq{0}.
$$
We study asymptotic behavior of $g_{n,p}(z)$ in the limit
$n,p\rightarrow\infty$, for $z\in\Lambda_{\eta}$,
$$
\Lambda_{v} =\{z\in\mathbf{ C}:
 \   |\I z|\geq{2v+1}\}.
\eqno (2.4)
$$

Our first statement generalizes the result about the semicircle law
in dilute Wigner ensemble of random matrices obtained under more restrictive conditions
\cite{KKPS}.
 \vskip0,5cm

\begin{theo} If the family of random variables ${\cal A}_n$ (2.1) is such that $\mu_{2+\rho}<\infty$ with $\rho>0$, then $g_{n,p}(z)$ determined by (2.2) and (2.3) converges in probability:
$$
P-\lim_{n,p\rightarrow\infty}g_{n,p}(z)=w(z), \
 \quad z\in\Lambda_{v},
 \eqno (2.5)
$$
where the   function $w(z)$ verifies equation
$$
w(z)=\frac{1}{-z-v^{2} w(z)}, \quad \I z\neq 0;
\eqno (2.6)
$$
$w(z)$ uniquely determines the semicircle distribution (1.3) being its Stieltjes transform
and therefore  (2.5) implies the weak convergence in probability
$$
\sigma(\lambda; H_{n,p})\to \s_{sc}(\lambda), \quad n,p\to\infty.
$$
\end{theo}

 \vskip0,2cm

The proof of Theorem 2.1 is based on the following two
asymptotic relations:
$$
\lim_{n,p\rightarrow\infty}{\bf E}\{g_{n,p}(z)\}=w(z), \
 \quad z\in\Lambda_{v}
\eqno (2.7)
$$
and
$$
{\bf Var}\{g_{n,p}(z)\}=o(1), \ \quad z\in\Lambda_{v}, \quad as \
n,p\rightarrow\infty.
\eqno (2.8)
$$
Indeed, convergence (2.5) can be deduced from (2.7) and (2.8) with the help of the standard arguments
(see for example \cite{A} or \cite{KP}).

\vskip0,3cm

 The further improvement of (2.8) is related with the
asymptotic properties of the resolvent covariance function
$$
C_{n,p}(z_{1},z_{2})={\bf E}\{g_{n,p}(z_{1})g_{n,p}(z_{2})\}-{\bf
E}\{g_{n,p}(z_{1})\}{\bf E}\{g_{n,p}(z_{2})\}.
$$
Let us formulate corresponding statement.

\vskip0,5cm

\begin{theo} Let ${\cal A}_{n}$ be such that, in addition
to (2.1), the following properties are verified :
$$
{\bf E}\{a(i,j)^{3}\}={\bf E}\{a(i,j)^{5}\}=0, \quad {\bf
E}\{a(i,j)^{4}\}=V_{4}(1+\delta_{ij})^{2},
\eqno (2.9)
$$
and $\mu_{14}<\infty$. Then in the limit
$n,p\rightarrow\infty$ such that
$$
p=O(n^{\alpha}),  \quad 3/5<\alpha\le 1,
\eqno (2.10)
$$
 equality
$$
C_{n,p}(z_{1},z_{2}) =
\frac{2v^2}{n^{2}}S(z_{1},z_{2}) +
\left( {2V_4\over np} - {6v^4\over n^2}\right) T(z_{1},z_{2}) + o(n^{-2})
\eqno (2.11)
$$
holds for all $z_{l}\in\Lambda_{v}$ with $S$ and $T$ given by the formulas
$$
S(z_{1},z_{2})=
\frac{1}{(1-v^{2}w_{1}^{2})( 1-
v^{2}w_{2}^{2})}\left(\frac{w_{1}-w_{2}}{z_{1}-z_{2}}\right)^{2},
\eqno (2.12)
$$
and
$$
T(z_{1},z_{2})=
\frac{{ w_{1}^{3}w_{2}^{3}}}{(1 -
v^{2}w_{1}^{2})(1 - v^{2}w_{2}^{2})}\ ,
\eqno (2.13)
$$
where $w_{1}=w(z_{1})$ and $w_{2}=w(z_{2})$ are  the solutions of (2.6).
\end{theo}

\vskip0,3cm

Let us discuss  results of Theorem 2.2.
If one considers the particular case of (2.10) when  $p=n$, then (2.12) turns into
equality
$$
C_{n,p}(z_1,z_2) = {2\over n^2} \left( v^2 S(z_1,z_2) + K_4 T(z_1,z_2)\right) + o(n^{-2}),
\eqno (2.14)
$$
where $K_4= V_4-3v^4$ is the fourth cumulant of the random variable $a_{ij}$. Relation (2.14) coincides with that  derived in \cite{KKP} for the resolvent covariance function of the Wigner ensemble that one gets from $H_{n,p}$ (2.2) when taking $p=n$.
Therefore Theorem 2.2 generalizes the results of \cite{KKP}.

It was shown in \cite{KKP} that in the local scaling limit  in the bulk of the spectrum
$$
z_1= \lambda +{r\over 2n} + i0,\   z_2 = \lambda - {r\over 2 n}-i0 \quad
{\hbox {and}} \quad   \lambda\in (-2v,2v),
\eqno (2.15)
$$
the leading term of (2.14) converges to the expression (cf. (1.7))
$$
{2\over n^2} \left( v^2 S(z_1,z_2) + K_4 T(z_1,z_2)\right) \to - {1\over r^2},
\eqno (2.16)
$$
where the  term with $K_4$ does not contribute.

In Section 6 we show that the leading term of $C_{n,p}(Z_1,z_2)$
(2.11) exhibits the same asymptotic behavior in the local scaling
limit (2.15) as the leading term of the resolvent covariance
function of the Wigner ensemble (2.16). This shows that the dilute
random matrices considered in the limit of the moderate dilution
(2.10) belong to the universality class of Wigner (non-diluted)
random matrices. The lower band $3/5$ in (2.10) is due to the
technical restrictions related with the cumulant expansions we use.
We discuss this question in more details at the end of the paper.
Pushing forward this order of the one could decrease the value of
the exponent $\alpha$ (2.10). But this demands under more computations
than of the present paper.

\newpage

\subsection{Cumulant expansions and resolvent identities}

We prove Theorem 2.1 and Theorem 2.2 by using the method proposed in
papers \cite{KKP} and \cite{KP} and further developed in
\cite{A,A2,KK}. The basic tools of this method are given by the
resolvent identities combined with the cumulant expansions
technique.
In present section we present these technical tools and explain
the scheme of the proofs of Theorems 2.1 and 2.2.

\subsubsection{The cumulant expansions formula} Let us consider a
family $\{X_{t}: \ t=1,\ldots,m\}$ of independent real random
variables defined on the same probability space such that ${\bf
E}\{|X_{t}|^{q+2}\}<\infty $ for some $q\in \N$ and $t=1,\ldots,m$.
Then for any complex-valued function $F(u_{1},\ldots,u_{m})$ of the
class $\mathcal{C}_{\infty}(\R^{m})$ and for all $j$, one has
$$
{\bf E}\{X_{t} F(X_{1},\ldots,X_{m})\}=\sum_{r=0}^{q}
\frac{K_{r+1}^{(X_t)}}{r!} {\bf
E}\left\{\frac{\partial^{r}F(X_{1},\ldots,X_{m})}{(\partial{X_{t}})^{r}}\right\}
  +  \epsilon_{q}(X_{t}),
\eqno (2.17)
$$
where $K_{r}^{(X_t)}=Cum_{r}(X_{t})$ is the r-th cumulant of $X_{t}$ and the
remainder $\epsilon_{q}(X_{t})$ can be estimated by inequality
$$
|\epsilon_{q}(X_{t})|\le{C_{q} \sup_{U\in \R^{m}}
|\frac{\partial^{q+1}F(U)}{\partial{u_{t}^{q+1}}}|{\bf
E}\{|X_{t}|^{q+2}\}},
\eqno (2.18)
$$
where $C_{q}$ is a constant. Relations (2.17) and
(2.18) can be proved by multiple using of the Taylor's
formula \cite{A} or by using the characteristic functions method (see for example, \cite{KKP}).

\vskip 0.3cm
{\it Remarks.}

1) The cumulants $K_{r}^{(X_t)}$ can be expressed in terms of the moments of $X-t$;
in particular,
$$
K_{1}^{(X_t)}=\breve{\mu}_{1}, \quad
K_{2}^{(X_t)}=\breve{\mu}_{2}-\breve{\mu}^{2}_{1}, \quad {\hbox{where }} \  \breve{\mu}_{r}={\bf E}(X_{t}^{r})
\eqno (2.19)
$$
Regarding the right-hand side of  (2.17) with $q=1$, we see that the remainder $\epsilon_{1}(X_{t})$ is given by the following relation
$$
\epsilon_{1}(X_{t})=-K_{2}^{(X_t)}{\bf
E}\left\{X_{t}\frac{\partial^{2}F(\tilde{x}^{(2)}_{t})}{\partial{X_{t}^{2}}}\right\}
-\frac{K_{1}^{(X_t)}}{2}{\bf
E}\left\{X^{2}_{t}\frac{\partial^{2}F(\tilde{x}^{(1)}_{t})}{\partial{X_{t}^{2}}}\right\}
$$
$$
+\frac{1}{2}{\bf
E}\left\{X_{t}^{3}\frac{\partial^{2}F(\tilde{x}^{(0)}_{t})}
{\partial{X_{t}^{2}}}\right\}. \eqno (2.20)
$$

2) In present paper we are mostly related with the case when ${\bf E}(X_{t})= {\bf
E}(X_{t}^{3})={\bf E}(X_{t}^{5})=0$. Then  $K_{1}^{(X_t)}=K_{3}^{(X_t)}=K_{5}^{(X_t)}=0$ and
$$
K_{2}^{(X_t)}=\breve{\mu}_{2}, \quad
K_{4}^{(X_t)}=\breve{\mu}_{4}-3\breve{\mu}^{2}_{2}, \quad
K_{6}^{(X_t)}=\breve{\mu}_{6}-15\breve{\mu}_{4}\breve{\mu}_{2}+30\breve{\mu}^{3}_{2}.
\eqno (2.21)
$$
In this case, the remainders $\epsilon_{q}(X_{t})$ of (2.17) considered  with $q=1$, $q=3$ and $q=5$ are as follows:
$$
\epsilon_{1}(X_{t})=-K_{2}^{(X_t)}{\bf
E}\left\{X_{t}\frac{\partial^{2}F(\tilde{x}^{(1)}_{t})}{\partial{X_{t}^{2}}}\right\}
+\frac{1}{2}{\bf
E}\left\{X_{t}^{3}\frac{\partial^{2}F(\tilde{x}^{(0)}_{t})}{\partial{X_{t}^{2}}}\right\},
\eqno (2.22)
$$
$$
\epsilon_{3}(X_{t})=-\frac{K_{4}^{(X_t)}}{3!}{\bf
E}\left\{X_{t}\frac{\partial^{4}F(\tilde{x}^{(2)}_{t})}
{\partial{X_{t}^{4}}}\right\} -\frac{K_{2}^{(X_t)}}{3!}{\bf
E}\left\{X_{t}^{3}\frac{\partial^{4}F(\tilde{x}^{(1)}_{t})}
{\partial{X_{t}^{4}}}\right\}
$$
$$ + \frac{1}{4!}{\bf
E}\left\{X_{t}^{5}\frac{\partial^{4}F(\tilde{x}^{(0)}_{t})}
{\partial{X_{t}^{4}}}\right\} \eqno (2.23)
$$
and
$$
\epsilon_{5}(X_{t})= - \frac{K_{6}^{(X_t)}}{5!}{\bf
E}\left\{X_{t}\frac{\partial^{6}F(\tilde{x}^{(3)}_{t})}
{\partial{X_{t}^{6}}}\right\} - \frac{K_{4}^{(X_t)}}{(3!)^{2}}{\bf
E}\left\{X_{t}^{3}\frac{\partial^{6}F(\tilde{x}^{(2)}_{t})}
{\partial{X_{t}^{6}}}\right\}
$$
$$
-\frac{K_{2}^{(X_t)}}{5!}{\bf
E}\left\{X_{t}^{5}\frac{\partial^{6}F(\tilde{x}^{(1)}_{t})}
{\partial{X_{t}^{6}}}\right\}
+ \frac{1}{6!}{\bf
E}\left\{X_{t}^{7}\frac{\partial^{6}F(\tilde{x}^{(0)}_{t})}
{\partial{X_{t}^{6}}}\right\},
\eqno (2.24)
$$
where for any given $\nu=0,\ldots,3$,  $\tilde{x}^{(\nu)}_{t}$ is a
real random variable that depends on $X_{t}$ and such that
$|\tilde{x}^{(\nu)}_{t}|\le{|X_{t}|}$. In what follows, we omit the superscripts ${(X_t)}$ in the cumulants and use the
following denotation
$$
{\partial^{r}{F(\tilde{x}_{t}^{(\nu)})}\over \partial{X_{t}^{r}}}=
\left[{\partial^{r}{F}\over \partial{X_{t}^{r}}}\right]^{(\nu)}.
$$

\vskip0,5cm

\subsubsection{Resolvent identities} Given two $n \times n$
matrices $A$ and $\tilde{A}$ such that $A^{-1}$ and $\tilde{A}^{-1}$
exist, we have
$$
{A}^{-1}=\tilde A^{-1}-\tilde A^{-1}(A-\tilde A)A^{-1}
\eqno (2.25)
$$
In the particular case, relation (2.25) leads to the resolvent identity
$$
\left( h-zI\right)^{-1}=\left(\tilde h-zI\right)^{-1}- \left(\tilde
h-zI\right)^{-1}\left( h-\tilde h\right)\left(h-zI\right)^{-1}
\eqno (2.26)
$$
is valid. Regarding (2.26) with $\tilde h=0$ and denoting
$G=\left(h-zI\right)^{-1}$, we get equality
$$
G(i,j) = \xi\delta_{ij} - \xi\sum_{s=1}^{n} G(i,s)h(s,j), \quad
\xi={-z^{-1}},
\eqno (2.27)
$$
where $h(i,j), \ i,j=1,\ldots,n$ are the entries of the matrix $h$,
$G(i,j)$ are the entries of the resolvent $G$ and $\delta$ denotes
the Kronecker symbol.

Using (2.26) we derive for $G=\left(h-zI\right)^{-1}$,
$|\I z|\neq{0}$ equality
$$
\frac{\partial{G(s,t)}}{\partial{h(j,k)}}=-\frac{1}{1+\delta_{jk}}\left[G(s,j)G(k,t)+
G(s,k)G(j,t)\right].
\eqno (2.28)
$$

We will also need two more formulas based on (2.28);
these are expressions for
$\partial^{2}{G(i,j)}/\partial{h(j,i)^{2}}$ and
$\partial^{3}{G(i,j)}/\partial^{3}{h(j,i)}$. We present them later.

\subsubsection{The scheme of the proof of the semicircle law}

 Let us explain the main idea of the proof of Theorem 2.1
that follows the lines of the paper \cite{KKP}. Here we consider the
case when $\mu_{3} = \sup_{i,j} \E  \vert a_{ij}\vert ^3<\infty$. Using (2.23) with
$h=H_{n,p}$ and denoting $\xi = -z^{-1}$, we can write that
$$
{\bf E}\{g_{n,p}(z)\}=\xi-\frac{\xi}{n}\sum_{i,j=1}^{n} {\bf
E}\{G_{n,p}(i,j)H_{n,p}(j,i)\}.
\eqno (2.29)
$$

To compute ${\bf E}\{G_{n,p}(i,j)H_{n,p}(j,i)\}$, we use relations
(2.17) and (2.28) and obtain the following
expressions (to simplify formulas we omit here and everywhere below
the subscripts $n$,$p$ when no confusion can arise);

\begin{itemize}
\item[$\bullet$] if $j<i$
$$
{\bf E}\{G(i,j)H(j,i)\}=K_{2}(j,i){\bf
E}\left\{\frac{\partial{G(i,j)}}{\partial{H(j,i)}}
\right\}+\epsilon^{(1)}_{ji}
$$
$$
=-\frac{v^{2}}{n}{\bf E}\{G(i,j)^{2}+
G(i,i)G(j,j)\}+\epsilon^{(1)}_{ji}
\eqno (2.30)
$$
with
$$
\epsilon^{(1)}_{ji}=K_{2}(j,i){\bf
E}\left\{H(j,i)\left[\frac{\partial^{2}{G(i,j)}}{\partial{H(j,i)^{2}}}\right]^{(1)}\right\}
+\frac{1}{2}{\bf
E}\left\{H(j,i)^{3}\left[\frac{\partial^{2}{G(i,j)}}{\partial{H(j,i)^{2}}}\right]^{(0)}\right\},
\eqno (2.31)
$$
where we used the denotations of the end of subsection 2.2.1.

In (2.30), we have used (2.28) in the form
$$
{\bf E
}\left\{\frac{\partial{G_{n,p}(i,j)}}{\partial{H_{n,p}(j,i)}}\right\}={\bf E
}\left\{\frac{\partial{G(i,j)}}{\partial{h(j,i)}}\vert_{h=H_{n,p}}\right\}.
$$

Also we have taken into account that
\begin{equation*}
K_{2}(j,i)=K_{2}\left(H_{n,p}(j,i)\right)=\frac{1}{p}{\bf
E}\{a(j,i)^{2}d_{n,p}(j,i)^{2}\}=\frac{v^{2}}{n}(1+\delta_{ji}).
\end{equation*}
\item[$\bullet$] If $i<j$, then using equality $H(j,i)=H(i,j)$, we
get
\begin{equation*}
{\bf E}\{G(i,j)H(i,j)\}=K_{2}(i,j){\bf
E}\left\{\frac{\partial{G(i,j)}}{\partial{H(i,j)}}
\right\}+\epsilon^{(2)}_{ij}
\end{equation*}
$$
=-\frac{v^{2}}{n}{\bf E}\{G(i,j)^{2}+
G(i,i)G(j,j)\}+\epsilon^{(2)}_{ij},
\eqno (2.32)
$$
where $\epsilon^{(2)}_{ij}$ is given by (2.31) with
$D_{ji}$ replaced by $D_{ij}$.

\item[$\bullet$] If $j=i$, then
\begin{equation*}
{\bf E}\{G(i,i)H(i,i)\}=K_{2}\left(H(i,i)\right){\bf
E}\left\{\frac{\partial{G(i,i)}}{\partial{H(i,i)}}
\right\}+\epsilon^{(3)}_{ii}
\end{equation*}
$$
=-\frac{2v^{2}}{n}{\bf E}\{G(i,i)^{2}\}+\epsilon^{(3)}_{ii},
\eqno (2.33)
$$
where $\epsilon^{(3)}_{ii}$ is given by (2.31) with
$D_{ji}$ replaced by $D_{ii}$.
\end{itemize}

Substituting (2.30), (2.32) and
(2.33)  into (2.29), we obtain equality
$$
{\bf E}\{g\}=\xi+\frac{\xi v^{2}}{n}\sum_{i,j=1}^{n}{\bf
E}\{G(i,j)^{2}+ G(i,i)G(j,j)\}+\epsilon
\eqno (2.34)
$$
with
\begin{equation*}
\epsilon=-\frac{\xi}{n}\sum_{i=1}^{n}\left[\sum_{j<i}\epsilon^{(1)}_{ji}
+\sum_{i<j}\epsilon^{(2)}_{ij}+ \epsilon^{(2)}_{ii}\right].
\end{equation*}
It is not hard to see that the terms $\epsilon^{(l)}_{ji}$, $l=1,2,3$, $z \in \Lambda_v$ (2.4)
are bounded by the same variable $2\mu_{3}(|\I z|^{3}n\sqrt{p})^{-1}$. We present the detailed computations in Section 3.
Then we can rewrite (2.34) in the form
$$
{\bf E}\{g_{n,p}(z)\}=\xi+\xi v^{2}{\bf
E}\{g^{2}_{n,p}(z)\}+\psi_{n,p}(z),
$$
where $\psi_{n,p}$ vanishes as $n,p\rightarrow\infty$ for all $z\in
\Lambda_{v}$.

Assuming that the average ${\bf E}\{\left(g_{n,p}(z)\right)^{2}\}$
factorizes ( see (2.8)), we obtain equality
$$
{\bf E}\{g_{n,p}(z)\}=\xi+\xi v^{2}{\bf
E}\{g_{n,p}(z)\}^{2}+\tilde{\psi}_{n,p}(z),
$$
where $\lim_{n,p\rightarrow\infty}\tilde{\psi}_{n,p}(z)=0$. Then one
can conclude that ${\bf
E}\{g_{n,p}(z)\})$ converges to  the solution of the
following equation (cf. (2.6) and (2.7)):
$$
w(z)=\xi + \xi v^{2}w(z)^{2}.
$$
 \vskip0,2cm

\subsubsection{The leading terms of resolvent covariance}
In this subsection we present the scheme  of the computation of the leading
terms of $C_{n,p}(z_{1},z_{2})$ (2.11).
Let us denote $g_{l}=g_{n,p}(z_{l}),\  l=1,2$. Given a
random variable, we consider its centered counterpart,  $f^{0}=f-{\bf E}f$. Using identity
$$
{\bf E}\{f^{0}g^{0}\}={\bf E}\{f^{0}g\},
\eqno (2.35)
$$
we rewrite $C_{12} =C_{n,p}(z_{1},z_{2})$ as
\begin{equation*}
C_{12}={\bf E}\{g^{0}_{1}g_{2}\}= \frac{1}{n}\sum_{i=1}^{n}{\bf
E}\{g^{0}_{1}G_{2}(i,i)\}.
\end{equation*}
Applying the resolvent identity (2.27) to $G_{2}(i,i)= G_{n,p}(i,j;z_2)$, we
obtain equality
$$
C_{12}=-\frac{\xi_{2}}{n}\sum_{i,j=1}^{n} {\bf
E}\{g^{0}_{1}G_{2}(i,j)H(j,i)\}.
\eqno (2.36)
$$
To compute ${\bf E}\{g^{0}_{1}G_{2}(i,j)H(j,i)\}$, we use again
(2.17) and get relation
\begin{equation*}
{\bf E}\{g^{0}_{1}G_{2}(i,j)H(j,i)\}=K_{2}{\bf
E}\left\{\frac{\partial\left(g^{0}_{1}G_{2}(i,j)\right)}{\partial{H(j,i)}}\right\}+\frac{K_{4}}{6}{\bf
E}\left\{\frac{\partial^{3}\left(g^{0}_{1}G_{2}(i,j)\right)}{\partial{H(j,i)^{3}}}\right\}+\tau_{ij}
\end{equation*}
where $K_{r}$ is the r-th cumulant of $H(j,i)$ and $\tau_{ij}$
vanishes.
Using twice
(2.28), we conclude that
\newpage
\begin{equation*}
\frac{\partial\{g^{0}_{1}G_{2}(i,j)\}}{\partial{H(j,i)}}=g^{0}_{1}\frac{\partial
G_{2}(i,j)}
{\partial{H(j,i)}}+G_{2}(i,j)\frac{1}{n}\sum_{s=1}^{n}\frac{\partial
G_{1}(s,s)}{\partial{H(j,i)}}
\end{equation*}
$$
=-g^{0}_{1}[G_{2}(i,j)^{2}+G_{2}(i,i)G_{2}(j,i)]-
\frac{2}{n}G^{2}_{1}(i,j)G_{2}(i,j).
\eqno (2.37)
$$
Then we get equality
\begin{equation*}
C_{12}=\xi_{2}v^{2}{\bf E}\{g^{0}_{1}g_{2}^{2}\}
+\frac{\xi_{2}v^{2}}{n^{2}}\sum_{i,j=1}^{n}{\bf
E}\{g^{0}_{1}G_{2}(i,j)^{2}\}+
\frac{2\xi_{2}v^{2}}{n^{3}}\sum_{i,j=1}^{n}{\bf
E}\{G_{1}^{2}(i,j)G_{2}(i,j)\}
\end{equation*}
$$
-\frac{\xi_{2}}{6n}\sum_{i,j=1}^{n}K_{4}{\bf
E}\left\{\frac{\partial^{3}\left(g^{0}_{1}G_{2}(i,j)\right)}
{\partial{H(j,i)^{3}}}\right\}+\Phi_{n,p}(z_{1},z_{2}),
\eqno (2.38)
$$
where $\Phi_{n,p}(z_{1},z_{2})$, $z_{l}\in \Lambda$ can be shown to vanish  in the
limit $n,p\rightarrow\infty$.

Regarding the right-hand side of (2.38), we apply to the third  term
the resolvent identity (2.25)
\begin{equation*}
G_{1}G_{2}=\frac{G_{1}-G_{2}}{z_{1}-z_{2}}
\end{equation*}
as follows;
$$
Tr G^{2}_{1}G_{2}=Tr G_{1}\frac{G_{1}-G_{2}}{z_{1}-z_{2}}=\frac{Tr
G^{2}_{1}}{z_{1}-z_{2}}-\frac{Tr G_{1}-Tr G_{2}}{(z_{1}-z_{2})^{2}}.
\eqno (2.39)
$$
Using identity (2.35) that gives equality
$$
{\bf E}\{g^{0}_{1}g_{2}^{2}\}=2{\bf E}\{g^{0}_{1}g_{2}\}{\bf
E}\{g_{2}\}+{\bf E}\{g^{0}_{1}(g^{0}_{2})^{2}\}
\eqno (2.40)
$$
 and taking into account (2.39), we rewrite
(2.38) in the form
\begin{equation*}
C_{12}=2\xi_{2}v^{2}{\bf E}\{g_{2}\}C_{12}+\frac{2\xi_{2}
v^{2}}{n^{2}}
\left[
\frac{1}{n}\frac{{\bf E}\{Tr
G^{2}_{1}\}}{z_{1}-z_{2}} -
\frac{{\bf E}\{g_{1}\}-{\bf
E}\{g_{2}\}}{(z_{1}-z_{2})^{2}}
\right]
\end{equation*}
$$
 -\frac{\xi_{2}}{6n}\sum_{i,j=1}^{n}K_{4}{\bf
E}\left\{\frac{\partial^{3}\left(g^{0}_{1}G_{2}(i,j)\right)}
{\partial{H(j,i)^{3}}}\right\}+\tilde{\Phi}_{n,p}(z_{1},z_{2}).
\eqno (2.41)
$$
Elementary transformations of the terms in brackets  based on
convergence (2.7) and equality (2.6) allows one to recognize the
terms $S(z_{1},z_{2})$ and $T(z_{1},z_{2})$ of (2.12) and (2.13)
that arise from the second and the third terms of the the right-hand
side of (2.41).


\section{Proof of Theorem 2.1}

Let us introduce a family of independent real random variables
${\cal \hat{A}}_{n,p}= \{\hat{a}_{p}(i,j): \ 1\le{i}\le{j}\le{n}\}$
defined by
\begin{equation*}
\hat{a}_{p}(i,j)=a(i,j)\mathbf{I}_{\{|a(i,j)|\le{\sqrt{p}}\}} =
\left\{
\begin{array}{lll}
a(i,j) & \textrm{if} & |a(i,j)|\le{\sqrt{p}} \\
0 &  \textrm{if} & |a(i,j)|>\sqrt{p}, \quad 0<p\le{n},
\end{array}\right.
\end{equation*}
where $\{a(i,j)\}$ verify conditions of Theorem 2.1. We define a
real symmetric $n\times n$ random matrix $\hat{H}_{n,p}$ by
equality:
\begin{equation*}
\hat{H}_{n,p}(i,j)=\frac{1}{\sqrt{p}}\hat{a}_{p}(i,j)d_{n,p}(i,j),
\quad 1\le{i}\le{j}\le{n}
\end{equation*}
and
consider the resolvent
$\hat{G}_{n,p}(z)=(\hat{H}_{n,p}-z)^{-1}, \quad \I z\neq{0}$.

Regarding
$\hat{g}_{n,p}(z)=n^{-1}\T \hat{G}_{n,p}(z)$, we prove in subsection 3.1 that
\begin{equation}\label{c.limnpEhatgnpz}
\lim_{n,p\rightarrow\infty}{\bf E}\{\hat{g}_{n,p}(z)\}=w(z) \
 \quad z\in\Lambda_{v}
\end{equation}
provided that the variance of $\hat g_{n,p}$ vanishes;

\begin{equation}\label{c.Varhatgnpz}
{\bf Var}\{\hat{g}_{n,p}(z)\}=o(1), \ \quad z\in\Lambda_{v}, \quad
\hbox{as} \ n,p\rightarrow\infty, \quad z\in \Lambda_{v}.
\end{equation}
We prove (3.2) in subsection 3.2.

At the end of this section   we show that
\begin{equation}\label{c.limnpEhatgnpz-gnpz}
\lim_{n,p\rightarrow\infty}{\bf E}|\hat{g}_{n,p}(z)-g_{n,p}(z)|=0,
\quad \hbox{as} \ n,p\rightarrow\infty, \quad z\in \Lambda_{v},
\end{equation}
where ${g}_{n,p}(z)$ is determined by (2.3).
Then relations (2.7) and (2.8) follow from (3.1) and (3.2) and  Theorem 2.1 is proved.

\vskip 0.3cm

The proofs of relations (3.1) and (3.2) represent the main subject of this section.
Let us start to perform this program.
The last general remark is that
in what follows, we will  use many times the following  two elementary inequalities
\begin{equation}\label{c.GijleImz}
|G(i,j)|\le{||G||}\le{\frac{1}{|\I z|}},
\end{equation}
and
\begin{equation}\label{c.sumGij2leImz2}
\sum_{j=1}^{n}|G(i,j)|^{2}=||G\vec{e}_{i}||^{2}\le{\frac{1}{|\I
z|^{2}}}, \quad i=1,\ldots,n
\end{equation}
that hold for the resolvent of any real symmetric matrix. Here and
below we consider $||e||_{2}^{2}=\sum_{i}|e(i)|^{2}$  and denote
by $||G||=\sup_{||e||_{2}=1}||Ge||_{2}$ the corresponding operator norm.\\

\subsection{Main relation for ${\bf E}\{\hat{g}_{n,p}(z)\}$}

Regarding (2.27) with $h=\hat{H}_{n,p}$, we can write that
\begin{equation}\label{c.Ehatgnpz}
{\bf E}\{\hat{g}_{n,p}(z)\}=\xi-\frac{\xi}{n}\sum_{i,j=1}^{n} {\bf
E}\{\hat{G}_{n,p}(i,j)\hat{H}_{n,p}(j,i)\}.
\end{equation}
To compute ${\bf E}\{\hat{G}_{n,p}(i,j)\hat{H}_{n,p}(j,i)\}$, we use
formula (2.17) with $q=1$ and equality (2.28) (everywhere
below, we omit the subscripts $n$,$p$ when no confusion can arise).
Then we get relation
\begin{equation}\label{c.EGijHji}
{\bf E}\{\hat{G}(i,j)\hat{H}(j,i)\}=\hat{K}_{1}(j,i){\bf
E}\{\hat{G}(i,j)\}+\hat{K}_{2}(j,i){\bf E}\left\{D_{ji}\hat{G}(i,j)
\right\}+\hat{\epsilon}_{ji}
\end{equation}
with
\begin{equation*}
\hat{\epsilon}_{ji}=-\hat{K}_{2}(j,i){\bf
E}\left\{\hat{H}(j,i)\left[D^{2}_{ji}\hat{G}(i,j)\right]^{(2)}
\right\}- \frac{\hat{K}_{1}(j,i)}{2}{\bf
E}\left\{\hat{H}(j,i)^{2}\left[D^{2}_{ji}\hat{G}(i,j)\right]^{(1)}
\right\}
\end{equation*}

\begin{equation*}
+\frac{1}{2}{\bf
E}\left\{\hat{H}(j,i)^{3}\left[D^{2}_{ji}\hat{G}(i,j)\right]^{(0)}\right\},
\end{equation*}
where we have denoted $D^{r}_{ji}=\partial^{r}/\partial{H(j,i)^{r}}$
and $\hat{K}_{r}(j,i)=Cum_{r}\left(\hat{H}_{n,p}(j,i)\right)$.

Substituting (3.7) into (3.6) and taking into account formula  (2.28),
 we obtain equality
\begin{equation}\label{c.Ehatgfinale}
{\bf E}\{\hat{g}\}=\xi+\frac{\xi v^{2}}{n^{2}}\sum_{i,j=1}^{n}{\bf
E}\{\hat{G}(i,j)^{2}+ \hat{G}(i,i)\hat{G}(j,j)\}+R,
\end{equation}
where
\begin{equation*}
R=-\frac{\xi}{n}\sum_{i,j=1}^{n}\hat{K}_{1}(j,i){\bf
E}\{\hat{G}(i,j)\}
\end{equation*}
\begin{equation}\label{c.R}
+\frac{\xi}{n^{2}}\sum_{i,j=1}^{n}[\hat{V}_{2}(j,i)-v^{2}]{\bf
E}\{\hat{G}(i,j)^{2}+
\hat{G}(i,i)\hat{G}(j,j)\}-\frac{\xi}{n}\sum_{i,j=1}^{n}\hat{\epsilon}_{ji}
\end{equation}
with $\hat{V}_{2}(j,i)=nK_{2}(j,i)/(1+\delta_{ji})$.

Now we can rewrite (\ref{c.Ehatgfinale}) in the
form
\begin{equation}\label{c.Eg}
{\bf E}\{\hat{g}\}=\xi+\xi v^{2}{\bf
E}\{\hat{g}\}^{2}+R+\phi_{1}+\phi_{2},
\end{equation}
where
\begin{equation}\label{c.phi1}
\phi_{1}=\frac{\xi v^{2}}{n^{2}}\sum_{i,j=1}^{n}{\bf
E}\{\hat{G}(i,j)^{2}\}
\end{equation}
and
\begin{equation}\label{c.phi2}
\phi_{2}=\xi v^{2}\left({\bf E}\{\hat{g}^{2}\}-{\bf
E}\{\hat{g}\}^{2}\right).
\end{equation}
Let us show that the terms $R$ and $\phi_{l}$, $l=1,2$, vanish in
the limit $n,p\rightarrow\infty$.

We start with $R$. Regarding the first term of the right-hand side  of
(\ref{c.R}) and using (\ref{c.GijleImz}), one obtains
\begin{equation*}
|\frac{\xi}{n}\sum_{i,j=1}^{n}\hat{K}_{1}(j,i){\bf
E}\{\hat{G}(i,j)\}|
\end{equation*}
\begin{equation}\label{c.Restimation1}
\le{\sum_{i,j=1}^{n}\frac{\sqrt{p}}{\eta n^{2}}{\bf
E}\left\{|a(j,i)|\mathbf{I}_{\{|a(j,i)|>\sqrt{p}\}}\frac{|a(j,i)|
^{1+\rho}}{\sqrt{p}^{1+\rho}}\right\}}\le{\frac{\mu_{2+\rho}}{\eta^{2}
p^{\rho/2}}}.
\end{equation}
To estimate the second term of the right-hand side  of  (\ref{c.R}), we
use (\ref{c.GijleImz}), and inequality
\begin{equation*}
|\hat{V}_{2}(j,i)-v^{2}|\le{|{\bf E}\{\hat{a}(j,i)^{2}\}-{\bf
E}\{a(j,i)^{2}\}|}
\end{equation*}
\begin{equation}\label{c.Restimation2}
\le{{\bf
E}\left\{\frac{|a(j,i)|^{2+\rho}}{p^{\rho/2}}\mathbf{I}_{\{|a(j,i)|>\sqrt{p}\}}\right\}}\le{
\frac{\mu_{2+\rho}}{p^{\rho/2}}}.
\end{equation}
Regarding the third term of the right-hand side of (\ref{c.R}) and
taking into account equality
\begin{equation}\label{c.D2jiGij}
D^{2}_{ji}\hat{G}(i,j)=\frac{1}{(1+\delta_{ji})^{2}}[2\hat{G}(i,j)^{3}+6\hat{G}(i,i)\hat{G}(j,j)
\hat{G}(i,j)],
\end{equation}
and estimate (\ref{c.GijleImz}), we conclude that
$|D^{2}_{ji}\hat{G}(i,j)| \leq{8|\I z|^{-3}}$ and that
\begin{equation}\label{c.Restimation3}
|\frac{\xi}{n}\sum_{i,j=1}^{n}\hat{\epsilon}_{ji}|\le{4\frac{|\xi|}{n}
\sum_{i,j=1}^{n} \left(\sup_{i,j}|D^{2}_{ji}\hat{G}(i,j)|\right){\bf
E}\{|\hat{H}(j,i)|^{3}\}}\le{\frac{24\mu_{2+\rho}}
{\eta^{4}p^{\rho/2}}}
\end{equation}

Now gathering relations given by (\ref{c.Restimation1}),
(\ref{c.Restimation2}) and (\ref{c.Restimation3}), we get  the
following bound for $R$:
\begin{equation}\label{c.Restimation}
|R|=O\left(\frac{1}{p^{\rho/2}}\right), \ as  \quad
n,p\rightarrow\infty.
\end{equation}

Inequality (\ref{c.sumGij2leImz2}) implies that
\begin{equation}\label{c.phi1estimation}
|\phi_{1}| \le{\frac{v^{2}}{\eta^{3}n}}, \quad \eta=2v+1.
\end{equation}

To estimate $\phi_{2}$ (\ref{c.phi2}), we use the elementary
inequality  $|{\bf E}\{\hat{g}^{2}\}-{\bf E}\{\hat{g}\}^{2}|\le{{\bf
Var}(\hat{g}_{n,p}(z))}$ and relation (\ref{c.Varhatgnpz}) that we
prove in the next subsection.

Relations (\ref{c.Varhatgnpz}), (\ref{c.Restimation})  and  (\ref{c.phi1estimation})
show that
\begin{equation}\label{c.phi1phi2phi3}
|R+\phi_{1}+\phi_{2}|=o(1), \ as \quad \ n,p\rightarrow\infty.
\end{equation}

Then equality (\ref{c.Eg}) and estimate (\ref{c.phi1phi2phi3})
imply that ${\bf E}\{\hat{g}_{n,p}(z)\}\rightarrow{w(z)}$, $z\in
\Lambda_{v}$, where $w(z)$ is the solution of equation
\begin{equation*}
w(z)=\xi+\xi v^{2}w(z)^{2},
\end{equation*}
such that $\I w(z) \cdot \I z>0$, $\I z\neq{0}$. This proves convergence in
average (\ref{c.Ehatgnpz}).


\subsection{Estimate of ${\bf Var}\{\hat{g}_{n,p}(z)\}$}

Let us denote $\hat{g}_{l}=n^{-1}\T \hat{G}_{n,p}(z_{l})$, $l=1,2$.
Then we can write relations (c.f. (2.36))
\begin{equation*}
{\bf E}\{\hat{g}^{0}_{1}\hat{g}^{0}_{2}\}={\bf
E}\{\hat{g}^{0}_{1}\hat{g}_{2}\}=-\frac{\xi_{2}}{n}\sum_{i,j=1}^{n}
{\bf E}\{\hat{g}^{0}_{1}\hat{G}_{2}(i,j)\hat{H}(j,i)\}.
\end{equation*}

For each pair $(i,j)$, $\hat{g}^{0}_{1}\hat{G}_{2}(i,j)$ is a smooth
function of $\hat{H}(j,i)$. Its derivatives are bounded because of
equation (2.28) and (\ref{c.GijleImz}). In particular,
$$
|D^{2}_{ji}\{\hat{g}^{0}_{1}\hat{G}_{2}(i,j)\}|
\leq{C\left(|\I z_{1}|^{-1} + |\I z_{2}|^{-1}\right)^{4} },
$$
 where $C$
is an absolute constant.

  According to the definition of $\hat{H}$ and the condition
$\mu^{2+\rho}<\infty$ of theorem, the third absolute moment of
$\hat{H}(j,i)$ is of order $1/(p^{\rho/2}n)$. Then we can apply
(2.17) with $q=1$ to ${\bf
E}\{\hat{g}^{0}_{1}\hat{G}_{2}(i,j)\hat{H}(j,i)\}$ and get relation

\begin{equation*}
{\bf
E}\{\hat{g}^{0}_{1}\hat{g}_{2}\}=-\frac{\xi_{2}}{n}\sum_{i,j=1}^{n}\hat{K}_{1}(j,i){\bf
E}\{\hat{g}^{0}_{1}\hat{G}_{2}{(i,j)}\}
\end{equation*}
\begin{equation}\label{c.Ehatg10g2finale}
-\frac{\xi_{2}}{n}\sum_{i,j=1}^{n}\hat{K}_{2}(j,i){\bf
E}\left\{D^{1}_{ji}\left(\hat{g}^{0}_{1}\hat{G}_{2}(i,j)\right)\right\}-
\frac{\xi_{2}}{n}\sum_{i,j=1}^ {n}\hat{\epsilon}^{'}_{ji},
\end{equation}
where $\hat{K}_{r}$ is the r-th cumulant of $\hat{H}(i,j)$ and
\begin{equation}\label{c.epsilon'ji}
|\hat{\epsilon}^{'}_{ji}|\le{4C\left(|\I z_{1}|^{-1} +
|\I z_{2}|^{-1}\right)^{4}{\bf
E}|\hat{H}(j,i)|^{3}}\le{\frac{64C\mu_{2+\rho}}{\eta^{4}np^{\rho/2}}}.
\end{equation}
Using expression  (2.37) and identity (2.40), we
rewrite (\ref{c.Ehatg10g2finale}) in the form
\begin{equation*}
{\bf E}\{\hat{g}^{0}_{1}\hat{g}_{2}\}=2\xi_{2}v^{2}{\bf
E}\{\hat{g}^{0}_{1}\hat{g}_{2}\}{\bf
E}\{\hat{g}_{2}\}+\xi_{2}v^{2}{\bf
E}\{\hat{g}^{0}_{1}(\hat{g}^{0}_{2})^{2}\}
\end{equation*}
\begin{equation*}
+\frac{\xi_{2}}{n^{2}}\sum_{i,j=1}^{n}V_{2}(j,i){\bf
E}\{\hat{g}_{1}^{0}\hat{G}_{2}(i,j)^{2}\}+R_{12},
\end{equation*}
where
\begin{equation*}
R_{12}= -\frac{\xi_{2}}{n}\sum_{i,j=1}^{n}\hat{K}_{1}(j,i){\bf
E}\{\hat{g}^{0}_{1}\hat{G}_{2}{(i,j)}\}+\frac{\xi_{2}}{n^{2}}
\sum_{i,j=1}^{n}[V_{2}(j,i)-v^{2}]{\bf
E}\{\hat{g}^{0}_{1}\hat{G}_{2}(i,i)\hat{G}_{2}(j,j)\}
\end{equation*}
\begin{equation}\label{c.R12}
\frac{2\xi_{2}}{n^{3}}\sum_{i,j=1}^{n}V_{2}(j,i){\bf
E}\{\hat{G}^{2}_{1}(i,j)\hat{G}_{2}(i,j)\}-\frac{\xi_{2}}{n}\sum_{i,j=1}^
{n}\hat{\epsilon}^{(1)}_{ij}.
\end{equation}
Introducing the   auxiliary variable
\begin{equation}\label{c.hatq2}
\hat{q}_{2}=\frac{\xi}{1-2\xi v^{2}{\bf E}\{\hat{g}_{2}\}},
\end{equation}
we can write the following relation
\begin{equation}\label{c.Ehatg10g2aestimer}
{\bf E}\{\hat{g}^{0}_{1}\hat{g}_{2}\}=\hat{q}_{2}v^{2}{\bf
E}\{\hat{g}^{0}_{1}(\hat{g}^{0}_{2})^{2}\}
+\frac{\hat{q}_{2}}{n^{2}}\sum_{i,j=1}^{n}V_{2}(j,i){\bf
E}\{\hat{g}_{1}^{0}\hat{G}_{2}(i,j)^{2}\}+
\frac{\hat{q}_{2}}{\xi_{2}}R_{12}.
\end{equation}
It is easy to see that
\begin{equation}\label{c.hatq2ineq}
|\hat{q}_{2}|\le{\frac{2}{|Imz_{2}|}} \quad z_{2}\in \Lambda_{v}.
\end{equation}
Then
\begin{equation*}
|\hat{q}_{2}v^{2}{\bf E}\{\hat{g}^{0}_{1}(\hat{g}^{0}_{2})^{2}\}
+\frac{\hat{q}_{2}}{n^{2}}\sum_{i,j=1}^{n}V_{2}(j,i){\bf
E}\{\hat{g}_{1}^{0}\hat{G}_{2}(i,j)^{2}\}|
\end{equation*}
\begin{equation}\label{c.Ehatg10g2aestimer1}
\le{\frac{4v^{2}}{\eta^{2}}\left({\bf Var}\{\hat{g}_{1}\}{\bf
Var}\{\hat{g}_{2}\}\right)^{1/2}+\frac{2\mu_{2+\rho}}{n\eta^{3}}\left({\bf
Var}\{\hat{g}_{1}\}\right)^{1/2}}.
\end{equation}

To estimate $R_{12}$ (\ref{c.R12}), we use inequality
\begin{equation}\label{c.R12estimation1}
\sum_{i=1}^{n}{\bf E}|\hat{G}^{b}_{1}(i,j)\hat{G}_{2}(i,j)|
\le{\left(\sum_{i=1}^{n}|\hat{G}^{b}_{1}(i,j)|^{2}\right)^ {1/2}
\left(\sum_{i=1}^{n}|\hat{G}_{2}(i,j)|^{2}\right)^{1/2}}\le{\frac{1}{\eta^{b+1}}}
\end{equation}
with $b=2$. Then computations similar to those of subsection 3.1
imply that
\begin{equation}\label{c.Ehatg10g2aestimer2}
|\frac{\hat{q}_{2}}{\xi_{2}}R_{12}|\le{
C\left[\frac{1}{p^{\rho/2}}+\frac{1}{n^{2}}\right]}
\end{equation}
where $C$ is a constant.

Considering (\ref{c.Ehatg10g2aestimer}) with $z=z_{1}=\bar{z}_{2}$
and using (\ref{c.Ehatg10g2aestimer1}) and
(\ref{c.Ehatg10g2aestimer2}), we get inequality
\begin{equation*}
{\bf Var}\{\hat{g}\}\le{ \frac{2v^{2}}{\eta^{2}}{\bf
Var}\{\hat{g}\}+\frac{4v^{2}}{n\eta^{3}}\sqrt{{\bf
Var}\{\hat{g}\}}+C\left[\frac{1}{p^{\rho/2}}+\frac{1}{n^{2}}\right]}.
\end{equation*}
Then (\ref{c.Varhatgnpz}) follows.


 Let us prove
(\ref{c.limnpEhatgnpz-gnpz}). Using the resolvent identity
(2.26), we can write that

\begin{equation*}
\hat{g}_{n,p}(z)-g_{n,p}(z)=\frac{1}{n}\sum_{i,s,t=1}^{n}\hat{G}(i,s)
[H-\hat{H}](s,t)G(t,i)=\frac{1}{n}\sum_{s,t=1}^{n}(G\hat{G})(s,t)
[H-\hat{H}](s,t)
\end{equation*}
This relation together with (\ref{c.GijleImz}) imply that
\begin{equation*}
{\bf E}|\hat{g}_{n,p}(z)-g_{n,p}(z)|\le{\frac{1}{\eta^{2}n\sqrt{p}}
\sum_{s,t=1}^{n}{\bf E}|a(s,t)-\hat{a}(s,t)|{\bf E}|d(s,t)|}
\end{equation*}
\begin{equation*}
\le{\frac{1}{\eta^{2}n\sqrt{p}} \sum_{s,t=1}^{n}\frac{p}{n}{\bf
E}\left\{|a(s,t)| \mathbf{I}_{\{|a(s,t)|>\sqrt{p}\}} \cdot
\frac{|a(s,t)|^{1+\rho}}{\sqrt{p}^{1+\rho}}\right\}}\le{\frac{\mu_{2+\rho}}{\eta^{2}p^{\rho/2}}}.
\end{equation*}
Then relation (\ref{c.limnpEhatgnpz-gnpz}) follows.


\section{Correlation Function of the Resolvent }    

In this section we give
 the  computations that represent the principal part of the proof of Theorem 2.2.
The auxiliary technical results will be proved in the next section.

\subsection{The scheme of the proof of Theorem 2.2}

Let us consider (2.36) and apply (2.17) to ${\bf
E}\{g^{0}_{1}G_{2}(i,j)H(j,i)\}$ with $q=5$. Taking into account
(2.9), we get relation
\begin{equation*}
C_{12}=-\frac{\xi_{2}}{n}\sum_{i,j=1}^{n}K_{2}{\bf
E}\left\{D^{1}_{ji}\left(g^{0}_{1}G_{2}(i,j)\right)\right\}-
\frac{\xi_{2}}{6n}\sum_{i,j=1}^{n}K_{4}{\bf
E}\left\{D^{3}_{ji}\left(g^{0}_{1}G_{2}(i,j)\right)\right\}
\end{equation*}
\begin{equation}\label{c.4C12cum}
-\frac{\xi_{2}}{120n}\sum_{i,j=1}^{n}K_{6}{\bf
E}\left\{D^{5}_{ji}\left(g^{0}_{1}G_{2}(i,j)\right)\right\}+\tau,
\end{equation}
where
\newpage
\begin{equation*}
\tau=\frac{\xi_{2}}{n}\sum_{i,j=1}^{n}\frac{K_{6}}{5!}{\bf
E}\left\{H(j,i)D^{6}_{pi}[g^{0}_{1}G_{2}(i,j)]^{(3)}\right\}
\end{equation*}
\begin{equation*}
+\frac{\xi_{2}}{n}\sum_{i,j=1}^{n}\frac{K_{4}}{(3!)^{2}}{\bf
E}\left\{H(j,i)^{3}D^{6}_{ji}[g^{0}_{1}G_{2}(i,j)]^{(2)}\right\}
\end{equation*}
\begin{equation*}
+\frac{\xi_{2}}{n}\sum_{i,j=1}^{n}\frac{K_{2}}{5!}{\bf
E}\left\{H(j,i)^{5}D^{6}_{ji}[g^{0}_{1}G_{2}(i,j)]^{(1)}\right\}
\end{equation*}
\begin{equation}\label{c.4tau}
-\frac{\xi_{2}}{n}\sum_{i,j=1}^{n}\frac{1}{6!}{\bf
E}\left\{H(j,i)^{7}D^{6}_{ji}[g^{0}_{1}G_{2}(i,j)]^{(0)}\right\}.
\end{equation}
Let us note that
\begin{equation}\label{c.4k2k4}
K_{2}=\frac{v^{2}}{n}(1+\delta_{ji}),\ \
K_{4}=\left(\frac{V_{4}}{np}
-\frac{3v^{4}}{n^{2}}\right)(1+\delta_{ji})^{2}=\frac{\Delta}{np}
(1+\delta_{ji})^{2}
\end{equation}
with
$$
\Delta=V_{4}-3v^{4}{p\over n}
$$
 and
\begin{equation}\label{c.4K6}
K_{6}=\left(\frac{V_{6}}{np^{2}}-\frac{15V_{4}v^{2}}
{n^{2}p}+\frac{30v^{6}}{n^{3}}\right)(1+\delta_{ji})^{3}
=\frac{\sigma}{np^{2}}(1+\delta_{ji})^{3}
\end{equation}
with $\sigma=V_{6}-15V_{4}v^{2}pn^{-1}+30v^{6}p^{2}n^{-2}$. In
(\ref{c.4tau}), we have denoted for each pair $(j,i)$

\begin{equation*}
[g^{0}_{1}G_{2}(i,j)]^{(\nu)}=\{g^{(\nu)}\}^{0}_{ji}(z_{1})
G^{(\nu)}_{ji}(i,j;z_{2}), \quad \nu=0,\ldots,3
\end{equation*}
and $G^{(\nu)}_{ji}(z_{l})=(H^{(\nu)}_{ji}-z_{l})^{-1}$, $l=1,2$
with real symmetric
\begin{equation*}
  H^{(\nu)}_{ji}(r,s)= \left\{
\begin{array}{lll}
H(r,s) & \textrm{if} & (r,s)\neq(j,i) \\
H^{(\nu)}(j,i) &  \textrm{if} & (r,s)=(j,i)
\end{array}\right.
\end{equation*}
where $|H^{(\nu)}(j,i)|\le{|H(j,i)|}$, $\nu=0,\ldots,3$.\\

Regarding the first term of the right-hand side of (\ref{c.4C12cum}), we can use
(2.37). Taking into account  (\ref{c.4k2k4}), we write that
\begin{equation*}
-\frac{\xi_{2}}{n}\sum_{i,j=1}^{n}K_{2}{\bf
E}\left\{D^{1}_{ji}\left(g^{0}_{1}G_{2}(i,j)\right)\right\}=
\xi_{2}v^{2}{\bf E}\{g^{0}_{1}g_{2}^{2}\}
+\frac{\xi_{2}v^{2}}{n^{2}}\sum_{i,j=1}^{n}{\bf
E}\{g^{0}_{1}G_{2}(i,j)^{2}\}
\end{equation*}
\begin{equation*}
+ \frac{2\xi_{2}v^{2}}{n^{3}}\sum_{i,j=1}^{n}{\bf
E}\{G_{1}^{2}(i,j)G_{2}(i,j)\}.
\end{equation*}
Using this equality and relations (2.39), (2.40) and (4.4),
 and computing the partial derivatives with the
help of (2.28), we get the following relation
\newpage
\begin{equation*}
C_{12}=2\xi_{2}v^{2}{\bf E}\{g_{2}\}C_{12}+\frac{2\xi_{2}
v^{2}}{n^{2}}\left[
-\frac{{\bf E}\{g_{1}\}-{\bf
E}\{g_{2}\}}{(z_{1}-z_{2})^{2}}+\frac{1}{n}\frac{{\bf E}\{\T
G_{1}^{2}\}}{z_{1}-z_{2}}\right]
\end{equation*}
\begin{equation}\label{c.4C12form1}
+\frac{2\xi_{2}\Delta}{n^{3}p}\sum_{i,j=1}^{n}{\bf
E}\{G_{1}^{2}(i,i)G_{1}(j,j)G_{2}(i,i)G_{2}(j,j)\}
+\sum_{r=1}^{8}Y_{r}+\Upsilon+\tau
\end{equation}
with
\begin{equation*}
Y_{1}=\xi_{2}v^{2}{\bf E}\{g^{0}_{1}(g_{2}^{0})^{2}\},
\end{equation*}
\begin{equation*}
Y_{2}=\frac{\xi_{2}v^{2}}{n}{\bf E}\left\{g^{0}_{1}\ \frac{1}{n}\T
G_{2}^{2}\right\},
\end{equation*}
\begin{equation*}
Y_{3}=\frac{\xi_{2}\Delta}{p}{\bf
E}\left\{g^{0}_{1}\left(\frac{1}{n}\sum_{i=1}^{n}G_{2}(i,i)^{2}
\right)^{2}\right\},
\end{equation*}
\begin{equation*}
Y_{4}=\frac{\xi_{2}\Delta}{n^{2}p}\sum_{i,j=1}^{n}\left({\bf
E}\{g^{0}_{1}G_{2}(i,j)^{4}\}+6{\bf
E}\{g^{0}_{1}G_{2}(i,j)^{2}G_{2}(i,i)G_{2}(j,j)\}\right),
\end{equation*}
\begin{equation*}
Y_{5}=\frac{2\xi_{2}\Delta}{n^{3}p}\sum_{i,j=1}^{n}{\bf
E}\{G_{1}^{2}(i,j)G_{2}(i,j)^{3}
+3G_{1}^{2}(i,j)G_{2}(i,j)G_{2}(i,i)G_{2}(j,j)\},
\end{equation*}
\begin{equation*}
Y_{6}=\frac{2\xi_{2}\Delta}{n^{3}p}\sum_{i,j=1}^{n}{\bf
E}\{G_{1}^{2}(i,j)G_{1}(i,j)G_{2}(i,j)^{2}+G_{1}^{2}(i,j)
G_{1}(i,j)G_{2}(i,i)G_{2}(j,j)\}
\end{equation*}
\begin{equation*}
+\frac{2\xi_{2}\Delta}{n^{3}p}\sum_{i,j=1}^{n}{\bf
E}\{G_{1}^{2}(i,i)G_{1}(j,j)G_{2}(i,j)^{2}\},
\end{equation*}
\begin{equation*}
Y_{7}=\frac{2\xi_{2}\Delta}{n^{3}p}\sum_{i,j=1}^{n}{\bf
E}\{G_{1}^{2}(i,j)G_{1}(i,j)^{2}G_{2}(i,j)+G_{1}^{2}(i,i)
G_{1}(j,j)G_{1}(i,j)G_{2}(i,j)\}
\end{equation*}
\begin{equation*}
+\frac{2\xi_{2}\Delta}{n^{3}p}\sum_{i,j=1}^{n}{\bf
E}\{G_{1}^{2}(i,j)G_{1}(j,j)G_{1}(i,i)G_{2}(i,j)\},
\end{equation*}
\begin{equation*}
Y_{8}=-\frac{3\xi_{2}\Delta}{n^{2}p}\sum_{i=1}^{n}\left({\bf
E}\{g_{1}^{0}G_{2}(i,i)^{4}\}+ \frac{1}{n}{\bf
E}\{G^{2}_{1}(i,i)G_{2}(i,i)^{3}\}\right)
\end{equation*}
\begin{equation*}
-\frac{3\xi_{2}\Delta}{n^{2}p}\sum_{i=1}^{n}\left(\frac{1}{n}{\bf
E}\{G^{2}_{1}(i,i)G_{1}(i,i)^{2}G_{2}(i,i)\}+\frac{1}{n}{\bf
E}\{G^{2}_{1}(i,i)G_{1}(i,i)G_{2}(i,i)^{2}\}\right)
\end{equation*}
and
\begin{equation*}
\Upsilon=-\frac{\xi_{2}}{120n}\sum_{i,j=1}^{n}\frac{\sigma
(1+\delta_{ji})^{3}}{np^{2}}{\bf
E}\left\{D^{5}_{ji}\left(g^{0}_{1}G_{2}(i,j)\right)\right\},
\end{equation*}
where $\tau$ is given by (\ref{c.4tau}).

Let us discuss the structure of relation (\ref{c.4C12form1}).
We see that the first term of the right-hand side
of (\ref{c.4C12form1}) is expressed in terms of $C_{12}$. This will
finally give a closed relation for $C_{12}$. The second and third
terms of the right-hand side  of (\ref{c.4C12form1}) give a non-zero
contribution to $C_{12}$ that provides the expressions of the
leading terms $S(z_{1},z_{2})$ (2.12) and  $T(z_{1},z_{2})$
(2.13). We
compute this contribution in subsection 4.3.

The three last
terms of (\ref{c.4C12form1}) contribute as the terms of the order  $o(n^{-2})$
in the limit (2.10). The following two statements give the detailed account on these vanishing terms.

\begin{lem} 
Under conditions of Theorem 2.2, the
estimates
\begin{equation}\label{c.4Y1}
|Y_{1}| =O\left(\{{\bf Var}(g_{1})\}^{1/2}[p^{-2}+{\bf
Var}(g_{2})]\right),
\end{equation}
\begin{equation}\label{c.4Y2}
|Y_{2}| =o\left(p^{-2}n^{-1}+p^{-2}\{{\bf
Var}(g_{1})\}^{1/2}+p^{-1}\{{\bf Var}(g_{1})\}^{1/2}\{{\bf
Var}(g_{2})\}^{1/2}\right)
\end{equation}
and
\begin{equation}\label{c.4Y3}
|Y_{3}| =O\left(p^{-2}\{{\bf Var}(g_{1})\}^{1/2}+p^{-1}\{{\bf
Var}(g_{1})\}^{1/2}\{{\bf Var}(g_{2})\}^{1/2}\right)
\end{equation}
are true in the limit $n,p\rightarrow\infty$.
\end{lem}

We postpone the proof of Lemma 4.1 to the next
section.

\vskip 0,4cm

\begin{lem} 
Under conditions of Theorem 2.2, the estimates
\begin{equation}\label{c.4maxYr}
\max_{r=4,\ldots,8}|Y_{r}| =O\left(n^{-1}p^{-1}[n^{-1}+\{{\bf
Var}(g_{1})\}^{1/2}]\right)
\end{equation}
\begin{equation}\label{c.4Upsilon}
|\Upsilon| =O\left(p^{-2}[n^{-1}+\{{\bf Var}(g_{1})\}^{1/2}]\right)
\end{equation}
\begin{equation}\label{c.4tauLemma}
|\tau| =o\left(n^{-1}p^{-1}[n^{-1}+\{{\bf
Var}(g_{1})\}^{1/2}]\right)
\end{equation}
are true in the limit $n,p\rightarrow\infty$.
\end{lem}

{\it Proof of Lemma 4.2.} We start with (\ref{c.4maxYr}). Inequality
(\ref{c.GijleImz}) and (\ref{c.sumGij2leImz2}) imply that if
$z_{l}\in\Lambda_{v}$, then
\begin{equation*}
|Y_{4}|\le{\frac{7\Delta}{\eta^{3}n^{2}p}\sum_{i,j=1}^{n}{\bf
E}|g_{1}^{0}G_{2}(i,j)^{2}|}=O\left(\frac{1}{np}\{{\bf
Var}(g_{1})\}^{1/2}\right).
\end{equation*}
Inequality (\ref{c.GijleImz}) and (\ref{c.R12estimation1}) with
$b=2$ imply inequality  $|Y_{5}|\le{8c\Delta/(\eta^{3}n^{2}p)}$. Using
(\ref{c.GijleImz}), (\ref{c.sumGij2leImz2}) and
(\ref{c.R12estimation1}) with $b=1$, we obtain that the terms
$Y_{6}$, $Y_{7}$ and $Y_{8}$ are all of the order indicated in
(\ref{c.4maxYr}).

\vskip0,2cm
Regarding   (\ref{c.4Upsilon}), it is not hard to see that this result follows from  the estimate
\begin{equation}\label{c.4ED5jig10G2ij}
{\bf E}|D^{5}_{ji}\{g^{0}_{1}G_{2}(i,j)\}|=O\left(n^{-1}+\{{\bf
Var}(g_{1})\}^{1/2}\right)
\end{equation}
 in the limit $n,p\to\infty$ (2.10).
Let us prove (\ref{c.4ED5jig10G2ij}). Using (2.28)
and (\ref{c.GijleImz}), we get  relation
\begin{equation*}
D_{ji}\{g^{0}_{1}\}=\frac{1}{n}\sum_{t=1}^{n}D_{ji}\{G_{1}(t,t)\}
=-\frac{2}{n}G^{2}_{1}(i,j)=O\left(\frac{1}{n}\right)
\end{equation*}
for all $z_{1}\in\Lambda_{v}$.
It is easy to show that
\begin{equation}\label{c.Drjig10}
D^{r}_{ji}\{g^{0}_{1}\}=O\left(\frac{1}{n}\right), \quad
r=1,2,\ldots, \ z\in\Lambda_{v}.
\end{equation}
Then (\ref{c.4ED5jig10G2ij}) follows from (\ref{c.Drjig10}) and
(\ref{c.GijleImz}). Estimate (\ref{c.4Upsilon}) is
proved.\vskip0,2cm

To proceed with estimates of  $\tau$ (\ref{c.4tau}), then we use the
following simple statement. \vskip0,2cm

\begin{lem} 
\cite{A} If $z_{l}\in\Lambda_{\eta},
\ l=1,2$, under condition of Theorem 2.2, the estimates
\begin{equation}\label{c.4Vargnpnu}
{\bf Var}([g_{n,p}(z_{l})]^{(\nu)})=O\left({\bf
Var}(g_{n,p}(z_{l}))+p^{-1}n^{-2}\right), \quad \nu=0,\ldots,3
\end{equation}
and
\begin{equation}\label{c.4D6jig10G2ijnu}
D^{6}_{ji}\left([g^{0}_{1}G_{2}(i,j)]^{(\nu)}\right)=O\left(n^{-1}
|G^{(\nu)}_{2}(i,j)|+
|[g^{0}_{1}]^{(\nu)}G^{(\nu)}_{2}(i,j)|\right), \ \nu=0,\ldots,3
\end{equation}
are true in the limit  $n,b\longrightarrow \infty$.
\end{lem}

Lemma 4.3 is proved in \cite{A}. We do not present the details here.
\vskip0,2cm

Regarding the first term of the right-hand side of (\ref{c.4tau}) and using
(\ref{c.4Vargnpnu}) and (\ref{c.4D6jig10G2ijnu}), we obtain
inequality
\begin{equation*}
\sum_{j=1}^{n}K_{6}{\bf
E}|H(j,i)D^{6}_{ji}[g^{0}_{1}G_{2}(i,j)]^{(3)}|
\le{\sum_{j=1}^{n}\frac{8\sigma c}{np^{2}}\left(\frac{\hat{\mu}_{1}
p^{1/2}}{n^{2}}+\frac{\hat{\mu}^{1/2}_{2}}{n^{1/2}}\left({\bf
Var}([g_{1}] ^{(3)})\right)^{1/2}\right)}
\end{equation*}
\begin{equation}
\label{c.4taueatimation2finale}
=o\left(\frac{1}{np^{2}}+\frac{1}{p^{2}}\left({\bf
Var}(g_{1})\right)^{1/2}\right),
\end{equation}
where $c$ is a constant. Repeating previous computations of
(\ref{c.4taueatimation2finale}), we obtain that
\begin{equation*}
\sum_{j=1}^{n}K_{4}{\bf
E}|H(j,i)^{3}D^{6}_{ji}[g^{0}_{1}G_{2}(i,j)]^{(2)}|
+\sum_{j=1}^{n}K_{2}{\bf
E}|H(j,i)^{5}D^{6}_{ji}[g^{0}_{1}G_{2}(i,j)]^{(1)}|
\end{equation*}
\begin{equation}\label{c.4taueatimation3finale}
=o\left(\frac{1}{np^{2}}+ \frac{1}{p^{2}}\left({\bf
Var}(g_{1})\right)^{1/2}\right).
\end{equation}
Now, regarding the last term of (\ref{c.4tau}) and using
(4.15), we obtain inequality
\begin{equation*}
\frac{1}{n}\sum_{i,j=1}^{n}{\bf
E}|H(j,i)^{7}D^{6}_{ji}[g^{0}_{1}G_{2}(i,j)]^{(0)}|
\end{equation*}
\begin{equation}\label{c.4taueatimation1}
\le{\frac{c_1}{n}\sum_{i,j=1}^{n}\left({\bf
E}\frac{|H(j,i)|^{7}}{n}+{\bf
E}|H(j,i)^{7}[g^{0}_{1}]^{(0)}||G^{(0)}_{2}(i,j)|\right)},
\end{equation}
where $c_1$ is a constant.
Regarding the last term of(4.18) and using
(\ref{c.sumGij2leImz2}) and (\ref{c.4Vargnpnu}), we get
\newpage 
\begin{equation*}
\frac{1}{n}\sum_{i,j=1}^{n}{\bf
E}|H(j,i)|^{7}|[g_{1}^{0}]^{(0)}||G^{(0)}_{2}(i,j)|\le{
\frac{1}{n}\sum_{i,j=1}^{n} \sqrt{{\bf E}|H(i,j)|^{14}}\sqrt{{\bf
E}|[g^{0}]^{(0)}|^{2}|G^{(0)}_{2}(i,j)|^{2}}}
\end{equation*}
\begin{equation*}
\le{\frac{\hat{\mu}^{1/2}_{14}}{p^{3}n}\sum_{i=1}^{n}\sum_{j=1}^{n}
\frac{1}{\sqrt{n}}\sqrt{{\bf
E}|[g^{0}]^{(0)}||G^{(0)}_{2}(t,s)|^{2}} }
\le{\frac{\hat{\mu}^{1/2}_{14}}{p^{3}n}\sum_{i=1}^{n}\left(\sum_{j=1}^{n}
{\bf E}|[g^{0}]^{(0)}||G^{(0)}_{2}(t,s)|^{2}\right)^{1/2}
}
\end{equation*}
\begin{equation*}
\le{c_2\frac{\hat{\mu}^{1/2}_{14}}{\eta p^{3}}\left({\bf
Var}([g_{1}]^{(0)} )\right)^{1/2}}
\le{\frac{\hat{\mu}^{1/2}_{14}}{\eta p^{3}} [\left({\bf Var}(g_{1}
)\right)^{1/2}+\frac{1}{p^{1/2}n}]},
\end{equation*}
where $c_2$ is a constant.
Using this estimate and we rewrite (\ref{c.4taueatimation1}) in the
form
\begin{equation*}
\frac{1}{n}\sum_{i,j=1}^{n}{\bf
E}|H(j,i)^{7}D^{6}_{ji}[g^{0}_{1}G_{2}(i,j)]^{(0)}|
\le{\frac{c_{2}\hat{\mu}_{7}}{np^{5/2}}+c_1c_2\frac{\hat{\mu}^{1/2}_{14}}{\eta
p^{3}} \left(\left({\bf Var}(g_{1}
)\right)^{1/2}+\frac{1}{p^{1/2}n}\right)}
\end{equation*}
\begin{equation}\label{c.4taueatimation1finale}
=o\left(\frac{1}{np^{2}}+ \frac{1}{p^{2}}\left({\bf
Var}(g_{1})\right)^{1/2}\right),
\end{equation}
where $c$ is a constant. Then (\ref{c.4tauLemma}) follows from the
estimates given by relations (\ref{c.4taueatimation1finale}),
(\ref{c.4taueatimation2finale}) and (\ref{c.4taueatimation3finale}).
Lemma 4.2 is proved.

\vskip0,2cm

\subsection{Estimate of the variance}  
Using the definition of $q_{2}$ (\ref{c.hatq2}), we rewrite
(\ref{c.4C12form1}) in the form
\begin{equation}\label{c.4C12form2}
C_{12}=\frac{2v^{2}}{n^{2}}S_{n,p}+2\left(\frac{V_{4}}{np}-\frac{3v^{4}}{n^{2}}\right)T_{n,p}
+\frac{q_{2}}{\xi_{2}}\sum_{r=1}^{8}Y_{r} +\Upsilon+\tau,
\end{equation}
where
\begin{equation}\label{c.4Tnp}
S_{n,p}=q_{2}\left[- \frac{{\bf E}\{g_{1}\}-{\bf
E}\{g_{2}\}}{(z_{1}-z_{2})^{2}}+\frac{1}{n}\frac{{\bf E}\{Tr
G_{1}^{2}\}}{z_{1}-z_{2}}\right]
\end{equation}
and
\begin{equation}\label{c.4Qnp}
T_{n,p}=\frac{q_{2}}{n^{2}}\sum_{i,j=1}^{n}{\bf
E}\{G_{1}^{2}(i,i)G_{1}(j,j)G_{2}(i,i)G_{2}(j,j)\}.
\end{equation}
Using inequality (\ref{c.GijleImz}) and (\ref{c.hatq2ineq}), we
obtain that
\begin{equation}\label{c.4QSestimation}
\left|2\left(\frac{V_{4}}{np}-\frac{3v^{4}}{n^{2}}\right)T_{n,p}(z_{1},z_{2})\right|
\le{\frac{4V_{4}}{\eta^{6}np}+\frac{12v^{4}} {\eta^{6}n^{2}}}, \quad
z_{1},z_{2} \in\Lambda_{\eta}.
\end{equation}
Lemma 4.1 and Lemma 4.2 together with (\ref{c.hatq2ineq}) imply that
\begin{equation*}
\left|\frac{q_{2}}{\xi_{2}}\sum_{r=1}^{8}Y_{r}+\Upsilon+\tau\right|\le{c\left(\frac{1}{np^{2}}+\frac{1}{n^{2}p}+\left[\frac{1}{p^{2}}
+\frac{1}{np}\right]\{{\bf Var}(g_{1})\}^{1/2}\right)}
\end{equation*}
\begin{equation}\label{c.4restestimation}
+c\left(\{{\bf Var}(g_{1})\}^{1/2}{\bf Var}(g_{2})+\frac{1}{p}\{{\bf
Var}(g_{1})\}^{1/2}\{{\bf Var}(g_{2})\}^{1/2}\right),
\end{equation}
where c is a constant. Using this inequality and relation
(2.10), (\ref{c.R12estimation1}) and
(\ref{c.4QSestimation}), we derive form (\ref{c.4C12form2}) the
following estimate
\begin{equation*}
{\bf Var}(g_{n,p}(z))\le{ \frac{A}{p^{2}}\sqrt{{\bf
Var}(g_{n,p}(z))}+\frac{B}{np}}
\end{equation*}
with A and B that depend on $z$ only. Since ${\bf Var}(g_{n,p}(z))$
is bounded for all $z\in \Lambda_{v}$, then we conclude that
\begin{equation}\label{c.4Vargnpz}
Var(g_{n,p}(z))=O\left(\frac{1}{np}\right).
\end{equation}
Substituting this estimate into (\ref{c.4restestimation}), we obtain
that
\begin{equation*}
\left|\frac{q_{2}}{\zeta_{2}}\sum_{r=1}^{8}Y_{r}+\Upsilon+\tau\right|=O\left(\frac{1}{p^{2}\sqrt{np}}\right).
\end{equation*}
This fact together with the restriction (2.10) implies that
\begin{equation*}
\frac{1}{p^{2}\sqrt{np}}\ll \frac{1}{n^{2}}
\end{equation*}
and that the estimate
\begin{equation}\label{c.4rest}
\left|\frac{q_{2}}{\zeta_{2}}\sum_{r=1}^{8}Y_{r}+\Upsilon+\tau\right|=o\left(\frac{1}{n^{2}}\right)
\end{equation}
holds. This proves (2.11).


\subsection{Leading terms of correlation function} 

To obtain the explicit expression for the leading term of
$C_{n,p}(z_{1},z_{2})$, it is necessary to study in detail the
variables $S_{n,p}$ and $T_{n,p}$. Let us formulate the
corresponding statements.
 \vskip0,2cm

\begin{lem}  If $z_{l}\in\Lambda_{v}, \ l=1,2$, then  under conditions of
Theorem 2.2,  the estimates
\begin{equation}\label{c.4EtrG12}
\frac{1}{n}{\bf E}{\hbox{Tr\,}} \, G_{l}^{2}= \frac{w^{2}_{l}}{1 -
v^{2}w^{2}_{l}}+O\left(\frac{1}{p}\right)
\end{equation}
 and
\begin{equation}\label{c.4EG12G1G2G1}
\frac{1}{n^{2}}\sum_{i,j=1}^{n}{\bf
E}\{G_{1}^{2}(i,i)G_{1}(j,j)G_{2}(i,i)G_{2}(j,j)\}=
\frac{w^{3}_{1}w_{2}^{2}}{1 -
v^{2}w^{2}_{1}}+O\left(\frac{1}{p}\right)
\end{equation}
hold in the limit $n,p\rightarrow\infty$ (2.10).
\end{lem}
\vskip0,2cm

{\it Proof of Lemma 4.4.} We start with (\ref{c.4EtrG12}) and
introduce the variable
\begin{equation*}
M(z)=M_{n,p}(z)=\frac{1}{n}\sum_{i,j=1}^{n}G(i,j)^{2}.
\end{equation*}
Applying identity (2.27) to $G(i,j)$ and using formula (2.17) with
$q=3$, we get relation
\begin{equation*}
{\bf E}\{M(z)\} = \xi{\bf E}\{g(z)\}+2\xi v^{2}{\bf
E}\{M(z)g(z)\}+\frac{2\xi v^{2}}{n^{2}}\sum_{i=1}^{n}G^{3}(i,i)
+\gamma_{0}
\end{equation*}
with
\begin{align}
\nonumber \gamma_{0}=&-\frac{\zeta}{6n}\sum_{j,s=1}^{n}K_{4}{\bf
E}\left\{D^{3}_{sj}(G^{2}(j,s))\right\}\\
\nonumber &-\frac{\zeta}{n4!}\sum_{j,s=1}^{n}{\bf
E}\left\{H(s,j)^{5}\left[D^{4}_{sj}(G^{2}(j,s))\right]^{(0)}\right\}\\
\nonumber &+\frac{\zeta}{n3!}\sum_{j,s=1}^{n}K_{2}{\bf
E}\left\{H(s,j)^{3}\left[D^{4}_{sj}(G^{2}(j,s))\right]^{(1)}\right\}\\
\label{c.4gamma0} &+\frac{\zeta}{n3!}\sum_{j,s=1}^{n}K_{4}{\bf
E}\left\{H(s,j)\left[D^{4}_{sj}(G^{2}(j,s))\right]^{(2)}\right\},
\end{align}
where $K_{r}$ are the cumulants of $H(s,j)$ (\ref{c.4k2k4}). Using
identity (c.f. (2.35))
\begin{equation}\label{c.4Efg}
{\bf E}\{fg\}={\bf E}\{fg^{0}\}+{\bf E}\{f\}{\bf E}\{g\},
\end{equation}
we obtain the following relation for ${\bf E}\{M(z)\}$:
\begin{equation}\label{c.4EMform1}
{\bf E}\{M(z)\} = \xi{\bf E}\{g(z)\}+2\xi v^{2}{\bf E}\{M(z)\}{\bf
E}\{g(z)\}+\gamma_{0}+\gamma_{1},
\end{equation}
where
\begin{equation*}
\gamma_{1} = 2\xi v^{2}{\bf E}\{M(z)g^{0}(z)\}+\frac{2\xi
v^{2}}{n^{2}}\sum_{i=1}^{n}G^{3}(i,i).
\end{equation*}
Relations (\ref{c.GijleImz}) and (\ref{c.4Vargnpz}) imply the
estimate
\begin{equation}\label{c.4gamma1estimation}
|\gamma_{1}|\le{\frac{2v^{2}}{\eta^{3}}
\left(\frac{1}{\sqrt{np}}+\frac{1}{n\eta}\right)}.
\end{equation}
Regarding relation (\ref{c.4gamma0}) and using (2.28) and
(\ref{c.GijleImz}), one obtains that
\begin{equation*}
\max_{r=3,4}\left(\sup_{s,j}|D^{r}_{sj} G^{2}(s,j)|\right)\le{c_{5}}
\end{equation*}
and that
\begin{equation}\label{c.4gamma0estimation}
|\gamma_{0}|\le{\frac{c_{5}nK_{4}}{6\eta}+\frac{|\zeta|}{n}\sum_{j,s=1}^{n}\sup_{s,j}|D^{4}_{sj}
G^{2}(s,j)|{\bf E}\{|H(s,j)|^{5}\}}\le{\frac{c_{5}nK_{4}}{6\eta}+
\frac{c_{5}\mu_{5}}{2\eta p^{3/2}}},
\end{equation}
where $c_{5}$ is a constant. Relations (\ref{c.4gamma1estimation})
and (\ref{c.4gamma0estimation}) imply that
\begin{equation}\label{c.4restEM}
|\gamma_{0}+\gamma_{1}|=O\left(\frac{1}{p}\right), \ \hbox{ as }
\quad n,p\rightarrow\infty.
\end{equation}
Using this estimate and (2.7), we derive from (\ref{c.4EMform1})
relation
\begin{equation*}
{\bf E}\{M(z)\}=\zeta w(z)[1-2\zeta
v^{2}w(z)]^{-1}+O\left(\frac{1}{p}\right), \quad z\in
\Lambda_{\eta}.
\end{equation*}
Then (\ref{c.4EtrG12}) follows from this relation and equality
(2.6). \vskip0,2cm

Now we prove (\ref{c.4EG12G1G2G1}). Let us consider  variable
\begin{equation*}
L(z_1,z_2)=\frac{1}{n^{2}}\sum_{i,j=1}^{n}{\bf
E}\{G_{1}^{2}(i,i)G_{1}(j,j)G_{2}(i,i)G_{2}(j,j)\}
\end{equation*}
Using (\ref{c.4Efg}), we obtain the following relation for $L(z)$
\begin{equation*}
 L(z_1,z_2) =\left(\frac{1}{n}\sum_{i=1}^{n}{\bf
E}\{G_{1}^{2}(i,i)G_{2}(i,i)\}\right)\left(\frac{1}{n}\sum_{j=1}^{n}{\bf
E}\{G_{1}(j,j)G_{2}(j,j)\}\right)\\
\end{equation*}
\begin{equation*}
 +\frac{1}{n}\sum_{i=1}^{n}{\bf
E}\{G_{1}^{2}(i,i)G_{2}(i,i)B^{0}_{12}\},
\end{equation*}
where
\begin{equation}
\label{B12}
B_{12}=\frac{1}{n}\sum_{j=1}^{n}{\bf E}\{G_{1}(j,j)G_{2}(j,j)\}.
\end{equation}
To proceed with the estimate of $L(z_1,z_2)$, we use the
following simple statement that we prove in the next section.
\vskip0,1cm

\begin{lem} If $z_{l}\in\Lambda_{\eta}, \ l=1,2$, under conditions of Theorem 5.2.2,
then the estimates
\begin{equation}\label{c.4VarB12}
{\bf Var}\{B_{12}\}=O\left([p^{-1}+({\bf Var}\{g_{1}\})^{1/2}+({\bf
Var}\{g_{2}\})^{1/2}]^{2}\right),
\end{equation}
\begin{equation}\label{c.4EB12}
{\bf E}\{B_{12}\}=w_{1}w_{2}+O\left(\frac{1}{p}\right)
\end{equation}
and
\begin{equation}\label{c.4EM12}
\frac{1}{n}\sum_{i=1}^{n}{\bf
E}\{G_{1}^{2}(i,i)G_{2}(i,i)\}=\frac{w^{2}_{1}w_{2}}{1-v^{2}w^{2}_{1}}+O\left(\frac{1}{p}\right)
\end{equation}
hold in the limit $n,p\rightarrow\infty$ (2.10).
\end{lem}
\vskip0,1cm

Now (\ref{c.4EG12G1G2G1}) follows from Lemma 4.5 and 
the definition of $L(z_1,z_2)$.
 Lemma 4.4 is proved.  $\hfill \blacksquare$
\vskip0,2cm

{\it Proof of Theorem 5.2.2.} Let us complete the proof of Theorem
2.2. It is easy to see that if $z_{2}\in\Lambda_{\eta}$, then the
definition of $q_{2}$ (\ref{c.hatq2}), the convergence (2.7) and
equation (2.6) imply that
\begin{equation}\label{c.4limq2}
\lim_{n,p\rightarrow\infty}q_{n,p}(z_{2})=\frac{w_{2}}{1-v^{2}w^{2}_{2}},
\quad z_{2}\in \Lambda_{\eta}.
\end{equation}
Finally, using Lemma 4.4 and relations (2.7) and
(\ref{c.4limq2}), we derive from (\ref{c.4C12form2}) relations
(2.11), (2.12) and (2.13). Theorem 2.2 is
proved. $\hfill \blacksquare$ \vskip0,2cm


\section{Proof of Auxiliary Statement} 

The main goal of this section is to prove Lemmas 4.1 and 4.5.
\vskip0,2cm

\subsection{Proof of Lemma 4.1}

\subsubsection{Estimate of $Y_{1}$ (\ref{c.4Y1})}

Variable $Y_{1}=\xi v^{2}{\bf E}\{g_{1}^{0}(g_{2}^{0})^{2}\}$
(\ref{c.4C12form1}) admits the obvious bound
\begin{equation}\label{c.5Y1ineg}
|Y_{1}|\le{\frac{v^{2}}{\eta}\sqrt{{\bf Var}\{g_{1}\}}\sqrt{{\bf
E}|g^{0}_{2}|^{4}}}.
\end{equation}
To proceed with (5.1), we prove the following
statement.

\vskip0,2cm

\begin{lem} If $z\in\Lambda_{v}$, then under conditions of Theorem 2.2,
the estimate
\begin{equation}\label{c.5Egnp0z4}
{\bf E}|g_{n,p}^{0}(z)|^{4}=O\left([p^{-2}+{\bf
Var}\{g_{n,p}(z)\}]^{2}\right)
\end{equation}
is true in the limit $n,p\rightarrow\infty$ (2.10).
\end{lem}
\vskip0,2cm

Let us note that the estimate (\ref{c.4Y1}) follows from
inequality (\ref{c.5Y1ineg}) and (\ref{c.5Egnp0z4}). \vskip0,2cm

{\it Proof of Lemma 5.1.} Let us consider the average
\begin{equation*}
W={\bf
E}\{g_{1}^{0}g_{2}^{0}g_{3}^{0}g_{4}^{0}\}=\frac{1}{n}\sum_{t=1}^{n}{\bf
E}T^{0}G_{4}(t,t)
\end{equation*}
with $T=g_{1}^{0}g_{2}^{0}g_{3}^{0}$. We apply to $G_{4}(t,t)$ the
resolvent identity (2.26) and obtain relation
\begin{equation*}
W=-\xi_{4}\sum_{t,s=1}^{n}{\bf E}\{T^{0}G_{4}(t,s)H(s,t)\}.
\end{equation*}
 Applying (2.17) with $q=3$  to ${\bf
E}\{T^{0}G_{4}(t,s)H(s,t)\}$ and taking into account
(2.28), we get relation
\begin{equation*}
W=\xi_{4}v^{2}{\bf E}\{T^{0}(g_{4})^{2}\}+\xi_{4}v^{2}{\bf
E}\{\frac{T^{0}}{n^{2}}\sum_{t,s=1}^{n}G_{4}(t,s)^{2}\}
\end{equation*}
\begin{equation*}
+ \frac{2\xi_{4}v^{2}}{n^{3}}\sum_{(i,j,k)}{\bf
E}\left(g_{i}^{0}g_{j}^{0}\sum_{t,y,s=1}^{n}G_{k}(y,s)G_{k}(t,y)G_{4}(t,s)\right)
\end{equation*}
\begin{equation}\label{c.5Wform1}
- \frac{\xi_{4}}{n}\sum_{t,s=1}^{n}\frac{K_{4}}{6}{\bf
E}\left\{D^{3}_{st}\left(T^{0}G_{4}(t,s)\right)\right\}+\Omega
\end{equation}
with
\begin{equation*}
\Omega=-\frac{\xi_{4}}{n4!}\sum_{t,s=1}^{n}{\bf
E}\left\{H(s,t)^{5}[D_{si}^{4}(T^{0}G_{4}(t,s))]^{(0)}\right\}+
\frac{\xi_{4}}{n}\sum_{t,s=1}^{n}K_{2}{\bf
E}\left\{H(s,t)^{3}[D_{si}^{4}(T^{0}G_{4}(t,s))]^{(1)}\right\}
\end{equation*}
\begin{equation}\label{c.5Omega}
+\frac{\xi_{4}}{n3!}\sum_{t,s=1}^{n}K_{4}{\bf
E}\left\{H(s,t)[D_{si}^{4}(T^{0}G_{4}(t,s))]^{(2)}\right\},
\end{equation}
where $K_{r}$ are the cumulants of $H(s,t)$ as in (\ref{c.4k2k4}).
In (\ref{c.5Wform1}), we introduce the notation
\begin{equation*}
\sum_{(i,j,k)}F(i,j,k)=F(1,2,3)+F(1,3,2)+F(2,3,1).
\end{equation*}
Applying to the first term of the RHS of (\ref{c.5Wform1}) relation
(2.39) and using the definition of $q_{4}$ (\ref{c.hatq2}), we
obtain that
\begin{equation*}
W=q_{4}v^{2}{\bf E}\{T^{0}(g^{0}_{4})^{2}\}+q_{4}v^{2}{\bf
E}\{\frac{T^{0}}{n^{2}}\sum_{t,s=1}^{n}G_{4}(t,s)^{2}\}
\end{equation*}
\begin{equation*}
+ \frac{2q_{4}v^{2}}{n^{3}}\sum_{(i,j,k)}{\bf
E}\left(g_{i}^{0}g_{j}^{0}\sum_{t,y,s=1}^{n}
G_{k}(y,s)G_{k}(t,y)G_{4}(t,s)\right)+\tilde{\Omega},
\end{equation*}
where
\begin{equation}\label{c.5tildOmega}
\tilde{\Omega}=-\frac{q_{4}}{n}\sum_{t,s=1}^{n}\frac{K_{4}}{6}{\bf
E}\left\{D^{3}_{st}(T^{0}G_{4}(t,s)
)\right\}+\frac{q_{4}}{\xi_{4}}\Omega.
\end{equation}
Regarding $G_{k}(t,\cdot)$ and $G_{4}(t,\cdot)$ in the third term of
the RHS of (\ref{c.Wformfinal}) as a vectors in $n$-dimensional
space, we derive from estimate (\ref{c.sumGij2leImz2}) that
\begin{align}
\nonumber &\left|\sum_{y,s=1}^{n}
G_{k}(y,s)G_{k}(t,y)G_{4}(t,s)\right|\\
\label{c.5estimGfg} &
\le{||G_{k}||\left(\sum_{y=1}^{n}|G_{k}(t,y)|^{2}\right)^{1/2}
\left(\sum_{s=1}^{n}|G_{4}(t,s)|^{2}\right)^{1/2}}\le{\frac{1}{\eta^{3}}}.
\end{align}
Now gathering relation given by (\ref{c.GijleImz}),
(\ref{c.sumGij2leImz2}),(\ref{c.hatq2ineq}), (\ref{c.5estimGfg}) and
\begin{equation*}
{\bf E}|T^{0}(g_{4}^{0})^{2}|\le{\frac{2}{\eta}[{\bf E}|T|{\bf
E}|g_{4}^{0}|+{\bf E}|Tg_{4}^{0}|]}
\end{equation*}
imply the following inequality for $W$:
\begin{equation}\label{c.5Wineg}
|W|\le{ \frac{4v^{2}}{\eta^{2}}{\bf
E}|Tg_{4}^{0}|+\frac{4v^{2}}{\eta^{2}}{\bf E}|T|{\bf
E}|g_{4}^{0}|+\frac{4v^{2}}{\eta^{3}n}{\bf E}|T|
+\frac{12v^{2}}{\eta^{4}n^{2}}{\bf E}|g_{i}^{0}g_{j}^{0}|
+|\tilde{\Omega}|}.
\end{equation}
Henceforth, for sake of clarity, we consider
$G=G_{1}=G_{3}=\bar{G}_{4}=\bar{G}_{2}$, then we get
$T=(g^{0})^{2}\bar{g}^{0}$ and
\begin{equation}\label{c.5ET}
{\bf E}|T|\le{\sqrt{{\bf E}|g^{0}|^{2}}\sqrt{{\bf
E}|g^{0}|^{4}}}=\sqrt{{\bf Var}\{g\}}\sqrt{{\bf Var}\{W\}}.
\end{equation}
Let us assume for the moment that
\begin{equation}\label{c.5tildOmegaestima}
|\tilde{\Omega}|=O\left(\frac{1}{pn^{3}}+\frac{1}{p^{3}n^{2}}+\frac{\sqrt{{\bf
Var}(g)}}{pn^{2}}+\frac{{\bf Var}(g)}{np}+\frac{\sqrt{{\bf
Var}(g)}\sqrt{W}}{p}+\frac{\sqrt{W}}{p^{2}}\right).
\end{equation}

Now returning to (\ref{c.5Wineg}) and gathering estimates given by
relations (\ref{c.GijleImz}), (\ref{c.5ET}) and
(\ref{c.5tildOmegaestima}) imply the following estimate
\begin{equation*}
W\le{A_{1}\left(\frac{1}{p}+\sqrt{{\bf
Var}(g)}\right)^{2}\sqrt{W}+\frac{A_{2}}{np}\left(\frac{1}{p}+\sqrt{{\bf
Var}(g)}\right)^{2}},
\end{equation*}
where $A_{1}$, $A_{2}$ are a constants. Then we obtain
(\ref{c.5Egnp0z4}).
\vskip0,2cm

To complete the proof of Lemma 5.1, let us prove
(\ref{c.5tildOmegaestima}). To do this, we use the following
statements.
\vskip0,5cm
\begin{lem} If $z\in\Lambda_{\eta}$, then under conditions
of Theorem 2.2 the estimates
\begin{equation}\label{c.5D3stT0Gts}
D^{3}_{st}\{T^{0}\bar{G}(t,s)\}=O\left(n^{-3}+n^{-2}|g^{0}|+n^{-1}|g^{0}|^{2}
+|T^{0}|\right),
\end{equation}
\begin{equation}\label{c.5D4stT0Gts}
[D^{4}_{st}(T^{0}\bar{G}(t,s))]^{(\nu)}=O\left(\left\{n^{-3}+n^{-2}|[g^{0}]
^{(\nu)}|+n^{-1}|[g^{0}]^{(\nu)}|^{2}
+|[T^{0}]^{(\nu)}|\right\}|G^{(\nu)}(t,s)|\right),
\end{equation}
and
\begin{equation}\label{c.5Eg0nur}
{\bf E}|[g^{0}]^{(\nu)}|^{r}=O\left(p^{-r/2}n^{-r}+{\bf
E}|g^{0}|^{r}\right),  \quad r=1,\ldots,4
\end{equation}
are true in the limit $n,b\to \infty$ satisfying
(2.10) and for all $\nu=0,1,2$.
\end{lem}
\vskip0,5cm

We prove this Lemma at the end of this subsection. \vskip0,2cm

Let us return to the proof of (\ref{c.5tildOmegaestima}). Regarding
the first term of the RHS of (\ref{c.5tildOmega}) and using the
definition of $K_{4}$ (\ref{c.4k2k4}), inequality
(\ref{c.hatq2ineq}) and estimate (\ref{c.5D3stT0Gts}), one gets with
the help of (\ref{c.5ET}) that
\begin{equation*}
|\frac{q_{4}}{n}\sum_{t,s=1}^{n}\frac{2\Delta}{3np}{\bf
E}\left\{D^{3}_{st}(T^{0}\bar{G}(t,s))\right\}|
\end{equation*}
\begin{equation}\label{c.5tildOmegaestima1}
=O\left(\frac{1}{pn^{3}}+\frac{\sqrt{{\bf
Var}(g)}}{pn^{2}}+\frac{{\bf Var}(g)}{np}+\frac{\sqrt{{\bf
Var}(g)}\sqrt{W}}{p}\right).
\end{equation}
Now let us estimate $\Omega$ (\ref{c.5Omega}). Regarding the first
term of the RHS of (\ref{c.5Omega}) and using (\ref{c.5D4stT0Gts}),
we obtain inequality
\begin{equation*}
\frac{1}{n}\sum_{t,s=1}^{n}{\bf
E}|H(s,t)^{5}[D^{4}_{st}(T^{0}\bar{G}(s,t))]^{(0)}|
\end{equation*}
\begin{equation*}
\le{\frac{c}{n}\sum_{t,s=1}^{n}{\bf
E}\left(\frac{|H(s,t)|^{5}}{n^{3}}+\frac{|H(s,t)^{5}[g^{0}]^{(0)}|}
{n^{2}}+ \frac{|H(s,t)^{5}||[g^{0}]^{(0)}|^{2}}{n}+|H(s,t)^{5}|{\bf
E}|[T]^{(0)}|\right)}
\end{equation*}
\begin{equation*}
+\frac{c}{n}\sum_{t,s=1}^{n}{\bf
E}|H(s,t)^{5}||[g^{0}]^{(0)}|^{3}|G^{(0)}(t,s)|.
\end{equation*}
To estimate the last term of this inequality, we use
(\ref{c.sumGij2leImz2}) and (\ref{c.5Eg0nur}), and we get estimate
\begin{equation*}
\frac{1}{n}\sum_{t,s=1}^{n}{\bf
E}|H(s,t)^{5}||[g^{0}]^{(0)}|^{3}|G^{(0)}(t,s)|\le{\frac{1}{n}\sum_{t,s=1}^{n}
\sqrt{{\bf E}|H(s,t)|^{10}}\sqrt{{\bf
E}|[g^{0}]^{(0)}|^{6}|G^{(0)}(t,s)|^{2}}}
\end{equation*}
\begin{equation*}
\le{\frac{\mu^{1/2}_{10}}{p^{2}n}\sum_{t=1}^{n}\left(\sum_{s=1}^{n}
{\bf E}|[g^{0}]^{(0)}|^{6}|G^{(0)}(t,s)|^{2}\right)^{1/2}
}=O\left(\frac{1}{p^{2}}\sqrt{W^{(0)}}\right)=O\left(\frac{1}{p^{2}}
[\sqrt{W}+\frac{1}{pn^{2}}]\right).
\end{equation*}
Using this estimate, relation (\ref{c.5Eg0nur}) and the same
arguments in the proof of  estimate of (\ref{c.4tauLemma}), one
obtains that
\begin{equation*}
\frac{1}{n}\sum_{t,s=1}^{n} {\bf
E}|H(s,t)^{5}[D^{4}_{st}(T^{0}\bar{G}(s,t))]^{(0)}|
\end{equation*}
\begin{equation}\label{c.5Omegaestimation1}
=O\left(\frac{1}{p^{3/2}n^{3}}+\frac{1}{p^{3}n^{2}}+\frac{\sqrt{{\bf
Var}(g)}}{p^{2}n^{3/2}}+\frac{\sqrt{W}}{p^{2}}+\frac{\sqrt{{\bf
Var}(g)}\sqrt{W}}{p^{3/2}}\right).
\end{equation}
Repeating the arguments used to prove (\ref{c.5Omegaestimation1}),
it is easy to show that the term
\begin{equation*}
\frac{1}{n}\sum_{t,s=1}^{n}K_{4}{\bf
E}|H(s,t)[D^{4}_{st}(T^{0}\bar{G}(s,t))]^{(1)}|+K_{2}{\bf
E}|H(s,t)^{3}[D^{4}_{st}(T^{0}\bar{G}(s,t))]^{(2)}|
\end{equation*}
is of the order indicated in the RHS in (\ref{c.5Omegaestimation1})
and that
\begin{equation}\label{c.5Omegaestimation}
\Omega=O\left(\frac{1}{p^{3/2}n^{3}}+\frac{1}{p^{3}n^{2}}+\frac{\sqrt{{\bf
Var}(g)}}{p^{2}n^{3/2}}+\frac{\sqrt{W}}{p^{2}}+\frac{\sqrt{{\bf
Var}(g)}\sqrt{W}}{p^{3/2}}\right).
\end{equation}

Then the estimate (\ref{c.5tildOmegaestima}) follows from
(\ref{c.5tildOmegaestima1}) and (\ref{c.5Omegaestimation}). Lemma
5.1 is proved.
\vskip0,2cm

{\it Proof of Lemma 5.2.}
We start with (\ref{c.5D3stT0Gts}).
Remembering that $T=[g^{0}]^{2}\bar{g}^{0}$ and using
(2.28) and (\ref{c.Drjig10}), we obtain that
\begin{equation*}
{\bf E}|D^{1}_{st}\{ T^{0}\}|=O(n^{-1}|g^{0}|^{2}).
\end{equation*}
\begin{equation*}
D^{2}_{st}\{T^{0}\}=O\left(n^{-2}|g^{0}|+n^{-1}|g^{0}|^{2}\right),
\end{equation*}
\begin{equation*}
D^{3}_{st}\{T^{0}\}=O\left(n^{-3}+n^{-2}|g^{0}|+n^{-1}|g^{0}|^{2}
\right)
\end{equation*}
Now it is easy to show that (\ref{c.5D3stT0Gts}) is true.
\vskip0,2cm

Similar computations prove the estimate (\ref{c.5D4stT0Gts}).\\

Finally, let us prove (\ref{c.5Eg0nur}). To simplify computation,
we denote $[g]^{(\nu)}=\tilde{g}$. Then the resolvent identity
(2.26) implies that
\begin{equation*}
\tilde{g}=\frac{1}{n}\sum_{k=1}^{n}\tilde{G}(k,k)=\frac{1}{n}\sum_{k=1}^
{n}
G(k,k)-\frac{1}{n}\sum_{k,r,i=1}^{n}\tilde{G}(k,r)\{\tilde{H}-H\}(r,i)
G(i,k)
\end{equation*}
\begin{equation*}
=g-\frac{1}{n} {\hbox{Tr}}\, (G\tilde{G}\delta_{H})
\end{equation*}
with
\begin{equation*}
\delta_{H}(r,i)=\{\tilde{H}-H\}(r,i)=\left\{
\begin{array}{lll}
0 & \textrm{if} & (r,i)\neq(s,j) \\
\tilde{H}(s,j)-H(s,j) &  \textrm{if} & (r,i)=(s,j),
\end{array}\right.
\end{equation*}
where $0\le{|\tilde{H}(s,j)|}\le{|H(s,j)|}$. Then
\begin{equation}\label{c.5Etildgr}
{\bf E}|\tilde{g}^{0}|^{r}\le{ c{\bf E}|g^{0}|^{r}+
\frac{c}{n^{r}}{\bf E}|Tr (G\tilde{G}\delta_{H})-{\bf E}(Tr
(G\tilde{G}\delta_{H}))|^{r}}, \  r=1,\ldots,4,
\end{equation}
where $c$ is a constant. Using (\ref{c.GijleImz}), we obtain that
\begin{equation*}
{\bf E}|Tr (G\tilde{G}\delta_{H})-{\bf E}(Tr
(G\tilde{G}\delta_{H}))|\le{\frac{2}{\eta^{2}}{\bf E}\{|H(s,j)|+{\bf
E}(|H(s,j)|)\}}
\end{equation*}
\begin{equation}\label{c.5ETrGtildG}
\le{\frac{2}{\sqrt{p}\eta^{2}}\left({\bf E}\left\{|a(s,j)|+{\bf
E}|a(s,j)|\frac{p}{n}\right\}\frac{p}{n}+\left[{\bf
E}|a(s,j)|\frac{p}{n}\right]\left(1-\frac{p}{n}\right)\right)}=O\left(
\frac{1}{\sqrt{p}}\right).
\end{equation}
Relation (\ref{c.5Eg0nur}) follows from (\ref{c.5Etildgr}) and
estimate (\ref{c.5ETrGtildG}). Lemma 5.2 is proved. $\hfill
\blacksquare$ \vskip0,2cm


\subsubsection{Estimate of $Y_{2}$ (\ref{c.4Y2})} 

Remembering that $Y_{2}=\xi_{2}v^{2}n^{-2}\sum_{i,s}{\bf
E}\{g^{0}_{1}Tr G_{2}^{2}\}$, we consider the average
\begin{equation*}
\breve{Y}_{2}=\frac{1}{n^{2}}\sum_{i,s=1}^{n}{\bf
E}\{g^{0}_{1}G_{2}(i,s)^{2}\}
\end{equation*}
and apply to $G_{2}(i,s)$ the resolvent identity (2.26).
Then
\begin{equation*}
\breve{Y}_{2}=\frac{\xi_{2}}{n}C_{12}-\frac{\xi_{2}}{n^{2}}
\sum_{i,s,t=1}^{n}
{\bf E}\{g^{0}_{1}G_{2}(i,s)G_{2}(i,t)H(t,s)\}.
\end{equation*}
Applying (2.17) with $q=1$ to ${\bf
E}\{g^{0}_{1}G_{2}(i,s)G_{2}(i,t)H(t,s)\}$, we obtain that

\begin{equation}\label{c.5Y2nouv}
\breve{Y}_{2}=2\xi_{2}v^{2}{\bf
E}\{g_{2}\}\breve{Y}_{2}+\frac{\xi_{2}}{n}C_{12}+\frac{2\xi_{2}v^{2}}{n^{2}}\sum_{i,s=1}^{n}{\bf
E}\{g^{0}_{1}G_{2}(i,s)^{2}g_{2}^{0}\}+\sum_{r=1}^{3}\Theta_{r},
\end{equation}
where
\begin{equation*}
\Theta_{1}=\frac{2v^{2}\xi_{2}}{n^{4}}\sum_{s,t=1}^{n}{\bf
E}\{G^{2}_{1}(s,t)G^{2}_{2}(s,t)\}+\frac{2v^{2}\xi_{2}}{n^{3}}\sum_{s=1}^{n}{\bf
E}\{g^{0}_{1}G^{3}_{2}(s,s)\},
\end{equation*}
\begin{equation*}
\Theta_{2}=-\frac{\xi_{2}}{n^{2}}\sum_{t,s=1}^{n}\frac{K_{4}}{6}
{\bf E}\left\{D^{3}_{ts}(g^{0}_{1}G^{2}_{2}(t,s)) \right\}
\end{equation*}
and
\begin{equation*}
\Theta_{3}=-\frac{\xi_{4}}{n^{2}4!}\sum_{t,s=1}^{n}{\bf
E}\left\{H(t,s)^{5}[D_{ts}^{4}(g^{0}_{1}G^{2}_{2}(t,s))]^{(0)}\right\}+
\end{equation*}
\begin{equation*}
\frac{\xi_{4}}{n^{2}}\sum_{t,s=1}^{n}K_{2}{\bf
E}\left\{H(t,s)^{3}[D_{ts}^{4}(g^{0}_{1}G^{2}_{2}(t,s))]^{(1)}\right\}
\end{equation*}
\begin{equation*}
+\frac{\xi_{4}}{n^{2}3!}\sum_{t,s=1}^{n}K_{4}{\bf
E}\left\{H(t,s)[D_{ts}^{4}(g^{0}_{1}G^{2}_{2}(t,s))]^{(2)}\right\},
\end{equation*}
where $K_{r}$ are the cumulants of $H(t,s)$ as in (\ref{c.4k2k4}).

The term $2\xi_{2}v^{2}{\bf E}\{g_{2}\}\breve{Y}_{2}$ can be put to
the left-hand side of (\ref{c.5Y2nouv}). Using (\ref{c.GijleImz}),
(\ref{c.sumGij2leImz2}) and (\ref{c.R12estimation1}), it is easy to
show that the second and the third terms of the RHS of
(\ref{c.5Y2nouv}) and $\Theta_{1}$ are of the order indicated in the
RHS of (\ref{c.4Y2}). Using similar arguments as those of the proof
of (\ref{c.4Y1}) (see (4.16)-(4.19)) and the following estimate
(c.f.(\ref{c.4ED5jig10G2ij}))
\begin{equation*}
D_{ts}^{r}(g^{0}_{1}G^{2}_{2}(t,s))=O\left(n^{-1}+|g_{1}^{0}|\right),
\quad r=3,4,
\end{equation*}
we conclude that the terms $\Theta_{2}$ and $\Theta_{3}$ are of the
order indicated in the RHS of (\ref{c.4Y2}). Relation (\ref{c.4Y2})
is proved. $\hfill \blacksquare$ \vskip0,2cm

\subsubsection{Estimate of $Y_{3}$ (\ref{c.4Y3})}
We rewrite $Y_{3}$ in the form $Y_{3}=\zeta_{2}\Delta p^{-1}{\bf
E}\left\{g^{0}_{1}[B_{22}]^{2}\right\}$, where  (cf. (\ref{B12}))
\begin{equation*}
B_{22}=\frac{1}{n}\sum_{i=1}^{n}G_{2}(i,i)^{2}.
\end{equation*}
Let us note that the estimate (\ref{c.4Y3}) follows from (2.40),
inequality
\begin{equation*}
|{\bf E}\{g_{1}^{0}(B_{22})^{2}\}|\le{2{\bf
E}|g_{1}^{0}B_{22}^{0}|{\bf E}|B_{22}|+{\bf
E}|g_{1}^{0}(B_{22}^{0})^{2}| }\le{\frac{4}{\eta^{2}}\sqrt{{\bf
Var}\{g_{1}\}}\sqrt{{\bf Var}\{B_{22}\}}}
\end{equation*}
and estimate (\ref{c.4VarB12}). This proves (\ref{c.4Y3}). Lemma 4.1
is proved.

$\hfill \blacksquare$ \vskip0,2cm

\subsection{Proof of Lemma 4.5.}

\subsubsection{Estimate of ${\bf Var}\{B_{12}\}$ (\ref{c.4VarB12})}

Let us consider the average $\Pi=n^{-1}\sum_{j}{\bf
E}\{B_{12}^{0}\bar{G}_{1}(j,j)\bar{G}_{2}(j,j)\}$ and apply to
$\bar{G}_{2}(j,j)$ the resolvent identity (2.26), we obtain that
\begin{equation*}
\Pi=\bar{\zeta}_{2} {\bf
E}\{B_{12}^{0}\bar{g}_{1}\}-\frac{\bar{\zeta}_{2}}{n}\sum_{j,s=1}^{n}{\bf
E}\{B_{12}^{0}\bar{G}_{1}(j,j)\bar{G}_{2}(j,s)H(s,j)\}.
\end{equation*}
Now applying formulas (2.17) with $q=3$ and taking into account
relation (2.28), we obtain that
\begin{align}
\nonumber \Pi=&\bar{\zeta}_{2}{\bf
E}\{B_{12}^{0}\bar{g}_{1}\}+\bar{\zeta}_{2} v^{2}\Pi{\bf
E}\{\bar{g}_{2}\}+\bar{\zeta}_{2} v^{2}{\bf E}\{B_{12}^{0}\bar{B}_{12}\bar{g}_{2}^{0}\}\\
\nonumber &+\frac{\bar{\zeta}_{2} v^{2}}{n^{2}}\sum_{j,s=1}^{n}{\bf
E}\{B_{12}^{0}[\bar{G}_{1}(j,j)\bar{G}_{2}(j,s)^{2}+2\bar{G}_{1}(j,j)\bar{G}_{1}(j,s)\bar{G}_{2}(j,s)]\}\\
\nonumber &+\frac{2\bar{\zeta}_{2}
v^{2}}{n^{3}}\sum_{i,j,s=1}^{n}{\bf
E}\{G_{1}(i,i)G_{2}(i,s)G_{2}(i,j)\bar{G}_{1}(j,j)\bar{G}_{2}(j,s)\}\\
\label{c.5Pi} &+\frac{2\bar{\zeta}_{2}
v^{2}}{n^{3}}\sum_{i,j,s=1}^{n}{\bf
E}\{G_{1}(i,s)G_{1}(i,j)G_{2}(i,i)\bar{G}_{1}(j,j)\bar{G}_{2}(j,s)\}+U(z)
\end{align}
with
\begin{align}
\nonumber U(z)=&-\frac{\bar{\zeta}_{2}}{6n}\sum_{j,s=1}^{n}K_{4}{\bf
E}\left\{D_{sj}^{3}(B_{12}^{0}\bar{G}_{1}(j,j)\bar{G}_{2}(j,s))\right\}\\
\nonumber &-\frac{\bar{\zeta}_{2}}{n4!}\sum_{j,s=1}^{n}{\bf
E}\left\{H(s,j)^{5}[D_{sj}^{4}(B_{12}^{0}\bar{G}_{1}(j,j)\bar{G}_{2}(j,s))]^{(0)}
\right\}\\
\nonumber &+\frac{\bar{\zeta}_{2}}{n3!}\sum_{j,s=1}^{n}K_{2}{\bf
E}\left\{H(s,j)^{3}[D_{sj}^{4}(B_{12}^{0}\bar{G}_{1}(j,j)\bar{G}_{2}(j,s))]^{(1)}\right\}\\
\label{c.5Uz} &+\frac{\bar{\zeta}_{2}}{n3!}\sum_{j,s=1}^{n}K_{4}{\bf
E}\left\{H(s,j)[D_{sj}^{4}(B_{12}^{0}\bar{G}_{1}(j,j)\bar{G}_{2}(j,s))]^{(2)}\right\}
\end{align}
where $K_{r}$, $r=2,4$ are the cumulant of $H(s,j)$ as in
(\ref{c.4k2k4}). Let us assume for the moment that
\begin{equation}\label{c.5Uestim}
U(z)=O\left(\frac{1}{p^{2}}+\frac{1}{p}\sqrt{{\bf
Var}\{B_{12}\}}\right).
\end{equation}
Now returning to (\ref{c.5Pi}) and gathering estimates given by
relations (\ref{c.GijleImz}), (\ref{c.sumGij2leImz2}),
(\ref{c.5estimGfg}) and (\ref{c.5Uestim}) imply the following
estimate
\begin{equation*}
{\bf Var}\{B_{12}\}\le{A_{1}\left[\sqrt{{\bf
Var}\{g_{1}\}}+\sqrt{{\bf
Var}\{g_{2}\}}+\frac{1}{p}\right]\sqrt{{\bf
Var}\{B_{12}\}}+\frac{A_{2}}{p^{2}}},
\end{equation*}
where $A_{1}$ and $A_{2}$ are some constants. Then (\ref{c.4VarB12})
follows from this inequality. \vskip0,2cm

Now, let us prove (\ref{c.5Uestim}). To do this we use the following
statement. \vskip0,2cm

\begin{lem}
If $z\in\Lambda_{\eta}$, then under conditions of Theorem 5.2.2, the
estimates
\begin{equation}\label{c.5DsjB}
D^{r}_{sj}\{B_{12}^{0}\bar{G}_{1}(j,j)\bar{G}_{2}(j,s)\}=O\left(n^{-1}+|B_{12}^{0}|\right),
\quad r=3,4
\end{equation}
and
\begin{equation}\label{c.5VarBnu}
{\bf Var}\{[B_{12}]^{(\nu)}\}=O\left(p^{-1}n^{-2}+{\bf
Var}\{B_{12}\}\right)
\end{equation}
are true in the limit $n,p\rightarrow\infty$ satisfying (2.10) and
for all $\nu=0,1,2$.
\end{lem}
\vskip0,2cm

We prove this Lemma at the end of this subsection.

Let us return to the proof of (\ref{c.5Uestim}). Using the
definition $K_{4}$ (\ref{c.4k2k4}) and estimate (\ref{c.5DsjB}), it
is easy to show that the first term of the RHS of (\ref{c.5Uz}) is
of the order indicated in the RHS of (\ref{c.5Uestim}).

Regarding the first term of the RHS of (\ref{c.5Uz}) and using
(\ref{c.5DsjB}) and (\ref{c.5VarBnu}), we obtain inequality
\begin{align}
\nonumber &\frac{1}{n}\sum_{j,s=1}^{n}{\bf
E}\left|H(s,j)^{5}[D_{sj}^{4}(B_{12}^{0}\bar{G}_{1}(j,j)\bar{G}_{2}(j,s))]^{(0)}
\right|\\
\nonumber &\le{\frac{c}{n}\sum_{j,s=1}^{n}{\bf
E}\left\{\frac{|H(s,j)|^{5}}{n}+|H(s,j)|^{5}[B_{12}^{0}]^{(0)}\right\}}\\
\nonumber
&\le{\frac{c}{n}\sum_{j,s=1}^{n}\frac{\mu_{5}}{n^{2}p^{3/2}}
+\frac{c^{'}\mu_{10}^{1/2}}{p^{9/2}n^{1/2}}\left[\frac{1}{p^{1/2}n}+\sqrt{{\bf
Var}\{B_{12}\}}\right]}\\
\label{c.5Uestim1} &=O\left(\frac{1}{p^{2}}+\frac{1}{p}\sqrt{{\bf
Var}\{B_{12}\}}\right),
\end{align}
where $c$ and $c^{'}$ are some constants. Repeating the arguments
used to prove (\ref{c.5Uestim1}), it is easy to show that the third
and the fourth terms of the RHS of (\ref{c.5Uz}) are of the order
indicated in the RHS of (\ref{c.5Uestim}). This proves
(\ref{c.5Uestim}).

\vskip0,2cm
{\it Proof of Lemma 5.3.} The estimate (\ref{c.5DsjB}) follows from
(2.28), (\ref{c.GijleImz}), (\ref{c.sumGij2leImz2}) and
(\ref{c.R12estimation1}).

Let us prove (\ref{c.5VarBnu}). To simplify computation, we denote
$[G_{l}]^{(\nu)}=\tilde{G}_{l}$, $l=1,2$. Then the resolvent
identity (2.26) imply that
\begin{align*}
G_{l}(k,k)&=\tilde{G}_{l}(k,k)-\sum_{r,i=1}^{n}G_{l}(k,r)\delta_{H}(r,i)
\tilde{G}_{l}(i,k)\\
&=\tilde{G}_{l}(k,k)-G_{l}(k,s)[H(s,j)-\tilde{H}(s,j)]
\tilde{G}_{l}(j,k)
\end{align*}
for $l=1,2$ with
\begin{equation*}
\delta_{H}(r,i)=\{H-\tilde{H}\}(r,i)=\left\{
\begin{array}{lll}
0 & \textrm{if} & (r,i)\neq(s,j); \\
H(s,j)-\tilde{H}(s,j) &  \textrm{if} & (r,i)=(s,j),
\end{array}\right.
\end{equation*}
where $0\le{|\tilde{H}(s,j)|}\le{|H(s,j)|}$. Then
\begin{align*}
B_{12}&=\frac{1}{n}\sum_{k=1}^{n}G_{1}(k,k)G_{2}(k,k)\\
&=\tilde{B}_{12}
-\frac{1}{n}\sum_{k=1}^{n}G_{1}(k,s)G_{2}(k,s)\tilde{G}_{1}(j,k)\tilde{G}_{2}(j,k)\delta_{H}(s,j)^{2}\\
&-\frac{1}{n}\sum_{k=1}^{n}\tilde{G}_{1}(k,k)G_{2}(k,s)\tilde{G}_{2}(j,k)\delta_{H}(s,j)\\
&-\frac{1}{n}\sum_{k=1}^{n}\tilde{G}_{2}(k,k)G_{1}(k,s)\tilde{G}_{1}(j,k)\delta_{H}(s,j).
\end{align*}

It is not hard to see that the last equality together with  relations (\ref{c.GijleImz}) and
(\ref{c.R12estimation1}) implies that
$$
{\bf Var}\{\tilde{B}_{12}\} 
\le{4{\bf Var}\{B_{12}\}+\frac{c_{1}}{n^{2}}\left({\bf
E}|H(s,j)|^{4}+\left[{\bf E}|H(s,j)|^{2}\right]^{2}+
2{\bf
E}|H(s,j)|^{2}\right)}
$$
$$
\le{4{\bf Var}\{B_{12}\}+\frac{c}{n^{2}}\left(\frac{\mu_{4}}{np}
+\frac{\mu^{2}_{2}}{n^{2}}+\frac{2\mu_{2}}{n}\right)},
$$
where
where $c_1$ and $c$ are  constants. This proves (\ref{c.5VarBnu}). Lemma 5.3 is
proved and this proves of the estimate (\ref{c.4VarB12}). $\hfill
\blacksquare$ \vskip0,2cm

\subsubsection{Proof of relation (\ref{c.4EB12})}
Remembering that $B_{12}=n^{-1}\sum_{i}G_{1}(i,i)G_{2}(i,i)$.
Applying relation (2.26) to one of $G_{2}(i,i)$ and using formula
(2.17) with $q=3$, we get relation
\begin{equation*}
{\bf E}\{B_{12}\} = \zeta_{2}{\bf E}\{g_{1}\}+\zeta_{2} v^{2}{\bf
E}\{B_{12}g_{2}\}+\Gamma_{1}+\Gamma_{2}
\end{equation*}
with
\begin{align*}
\Gamma_{1}=&\frac{\zeta_{2}v^{2}}{n^{2}}\sum_{i,s=1}^{n}{\bf
E}\{G_{1}(i,s)G_{2}(i,s)G_{2}(i,i)+G_{1}(i,i)G_{2}(i,s)^{2}\}\\
&-\frac{\zeta_{2}}{6n} \sum_{i,s=1}^{n}K_{4}{\bf
E}\left\{D^{3}_{si}(G_{1}(i,i)G_{2}(i,s))\right\}\\
\Gamma_{2}=& -\frac{\zeta_{2}}{n}
\sum_{i,s=1}^{n}\tilde{\Gamma}_{is},
\end{align*}
where
\begin{equation*}
|\tilde{\Gamma}_{is}|\le{c\sup_{i,s}|D^{4}_{si}(G_{1}(i,i)G_{2}(i,s))|{\bf
E}|H(s,i)|^{5}}\le{\frac{c\mu_{5}}{\eta^{6}p^{3/2}n}},
\end{equation*}
$K_{r}$ are the cumulants of $H(s,i)$ as in (\ref{c.4k2k4}) and $c$
is a constant.

Using identity (\ref{c.4Efg}), we obtain the following relation for
${\bf E}\{B_{12}\}$:
\begin{equation}\label{c.5EB12final}
{\bf E}\{B_{12}\} = \frac{\zeta_{2}{\bf E}\{g_{1}\}}{1-\zeta_{2}{\bf
E}\{g_{2}\}}+\frac{1}{1-\zeta_{2}{\bf E}\{g_{2}\}}\left[\zeta_{2}
v^{2}{\bf E}\{B_{12}g^{0}_{2}\}+\Gamma_{1}+\Gamma_{2}\right]
\end{equation}
Relations (\ref{c.GijleImz}), (\ref{c.4Vargnpz}) and the estimate
\begin{equation*}
\max_{r=3,4}\left(\sup_{s,i}|D^{r}_{si}
(G_{1}(i,i)G_{2}(i,s))|\right)\le{c_{1}}
\end{equation*}
imply that
\begin{equation}\label{c.B12estimation1}
\left|\zeta_{2} v^{2}{\bf
E}\{B_{12}g^{0}_{2}\}+\Gamma_{1}+\Gamma_{2}\right|=O\left(\frac{1}{p}\right),
\end{equation}
where $c_{1}$ is a constant.

Now to proceed with the estimate of the first term of the RHS of
(\ref{c.5EB12final}), we use the following simple statement that we
prove in the end of this subsection. \vskip0,1cm

\begin{lem} If $z\in\Lambda_{\eta}$, under conditions of Theorem 2.2,
then the estimate
\begin{equation}\label{c.5Eg}
{\bf E}\{g_{n,p}(z)\}=w(z)+O\left(\frac{1}{p}\right)
\end{equation}
holds for enough $n$, $p$ satisfying (2.10).
\end{lem}
\vskip0,1cm

Using Lemma 5.4 and relation (\ref{c.B12estimation1}), we derive
from (\ref{c.5EB12final}) estimate (\ref{c.4EB12}).

\subsubsection{Proof of relation (\ref{c.4EM12})}

We introduce the variable
\begin{equation*}
U_{12}=\frac{1}{n}\sum_{i=1}^{n}G^{2}_{1}(i,i)G_{2}(i,i).
\end{equation*}
Applying relation (2.26) to one of $G_{2}(i,i)$ and using formula
(2.17) with $q=3$, we get relation
\begin{equation*}
{\bf E}\{U_{12}\} = \zeta_{2}\frac{1}{n}{\bf E}\{{\hbox{Tr\,}} \,
G^{2}_{1}\}+\zeta_{2} v^{2}{\bf E}\{U_{12}g_{2}\}+\Psi_{1}+\Psi_{2}
\end{equation*}
with
\begin{align*}
\Psi_{1}=&\frac{2\zeta_{2}v^{2}}{n^{2}}\sum_{i,s=1}^{n}{\bf
E}\{G^{2}_{1}(i,i)G_{1}(i,s)G_{2}(i,s)+G^{2}_{1}(i,s)G_{1}(i,i)G_{2}(i,s)\}\\
&+\frac{\zeta_{2}v^{2}}{n^{2}}\sum_{i,s=1}^{n}{\bf
E}\{G^{2}_{1}(i,i)G_{2}(i,s)^{2}\}-\frac{\zeta_{2}}{6n}
\sum_{i,s=1}^{n}K_{4}{\bf
E}\left\{D^{3}_{si}(G^{2}_{1}(i,i)G_{2}(i,s))\right\}\\
\Psi_{2}=& -\frac{\zeta_{2}}{n} \sum_{i,s=1}^{n}\tilde{\Psi}_{is},
\end{align*}
where
\begin{equation*}
|\tilde{\Psi}_{is}|\le{c\sup_{i,s}|D^{4}_{si}(G^{2}_{1}(i,i)G_{2}(i,s))|{\bf
E}|H(s,i)|^{5}}\le{\frac{c\mu_{5}}{\eta^{7}p^{3/2}n}},
\end{equation*}
$K_{r}$ are the cumulants of $H(s,i)$ as in (\ref{c.4k2k4}) and $c$
is a constant.

Using identity (\ref{c.4Efg}), we obtain the following relation for
${\bf E}\{U_{12}\}$:
\begin{equation}\label{c.5EU12final}
{\bf E}\{U_{12}\} = \frac{\zeta_{2}\frac{1}{n}{\bf E}\{{\hbox{Tr\,}}
\, G^{2}_{1}\}}{1-\zeta_{2}{\bf
E}\{g_{2}\}}+\frac{1}{1-\zeta_{2}{\bf E}\{g_{2}\}}\left[\zeta_{2}
v^{2}{\bf E}\{U_{12}g^{0}_{2}\}+\Psi_{1}+\Psi_{2}\right]
\end{equation}
Relations (\ref{c.GijleImz}), (\ref{c.4Vargnpz}) and the estimate
\begin{equation*}
\max_{r=3,4}\left(\sup_{s,i}|D^{r}_{si}
(G^{2}_{1}(i,i)G_{2}(i,s))|\right)\le{c_{1}}
\end{equation*}
imply that
\begin{equation}\label{c.U12estimation1}
\left|\zeta_{2} v^{2}{\bf
E}\{U_{12}g^{0}_{2}\}+\Gamma_{1}+\Gamma_{2}\right|=O\left(\frac{1}{p}\right),
\end{equation}
where $c_{1}$ is a constant.

Using relations (\ref{c.4EtrG12}), (\ref{c.U12estimation1}) and
(\ref{c.5Eg}), we derive from (\ref{c.5EU12final}) the estimate
(\ref{c.4EM12}). Lemma 4.5 is proved.

\vskip0,2cm

{\it Proof of Lemma 5.4}

Remembering that $g=n^{-1}\sum_{i}G(i,i)$. Applying relation (2.26)
to one of $G(i,i)$ and using formula (2.17) with $q=3$, we get
relation
\begin{equation}\label{c.5Egrelation}
{\bf E}\{g\} = \zeta +\zeta v^{2}{\bf E}\{g\}^{2}+\Phi_{1}+\Phi_{2}
\end{equation}
with
\begin{align*}
\Phi_{1}=&\frac{2\zeta v^{2}}{n^{2}}\sum_{i,s=1}^{n}{\bf
E}\{G(i,s)^{2}\}+\zeta v^{2}[{\bf E}\{g^{2}\}-{\bf E}\{g\}^{2}]\\
&-\frac{\zeta}{6n} \sum_{i,s=1}^{n}K_{4}{\bf
E}\left\{D^{3}_{si}(G(i,s))\right\}\\
\Phi_{2}=& -\frac{\zeta}{n} \sum_{i,s=1}^{n}\tilde{\Phi}_{is},
\end{align*}
where
\begin{equation*}
|\tilde{\Phi}_{is}|\le{c\sup_{i,s}|D^{4}_{si}(G(i,s))|{\bf
E}|H(s,i)|^{5}}\le{\frac{c\mu_{5}}{\eta^{5}p^{3/2}n}},
\end{equation*}
$K_{r}$ are the cumulants of $H(s,i)$ as in (\ref{c.4k2k4}) and $c$
is a constant.

Relations (\ref{c.GijleImz}), (\ref{c.4Vargnpz}) and the estimate
\begin{equation*}
\max_{r=3,4}\left(\sup_{s,i}|D^{r}_{si} (G(i,s))|\right)\le{c_{1}}
\end{equation*}
imply that
\begin{equation}\label{c.Egestimation1}
\left|\Phi_{1}+\Phi_{2}\right|=O\left(\frac{1}{p}\right),
\end{equation}

where $c_{1}$ is a constant. Using relation (\ref{c.Egestimation1}),
we derive from (\ref{c.5Egrelation}) the estimate (\ref{c.5Eg}).
Lemma 5.4 is proved. $\hfill \blacksquare$ \vskip0,2cm


\section{Scaling Limit And Universality Conjecture}   

 The asymptotic expression for $C_{n,p}(z_{1},z_{2})$ obtained in Theorem 2.2 and
$T(z_{1},z_{2})$ regarded in the limit
$z_{1}=\lambda_{1}+i\epsilon_{1}$, $z_{2}=\lambda_{2}+i\epsilon_{2}$
and $\epsilon_{j}\downarrow 0, j=1,2$ can supply the information
about the local properties of eigenvalue distribution. We follow the
schema proposed in \cite{KKP}.

Let us recall the inversion formula of the Stieltjes transform $w(z)$ (2.6) of the semicircle distribution
with the density $\rho_{sc} = \sigma_{sc}'$:
\begin{equation}\label{c.6roh}
\rho_{sc}(\lambda)=\pi^{-1}\lim_{\epsilon\downarrow 0}{\hbox{Im}}\,
w(\lambda+i\epsilon)=I_{\lambda}\{w(z)\}.
\end{equation}
Consider the density-density correlation function
\begin{equation}\label{c.6.2}
\digamma_{n,p}(\lambda_{1},\lambda_{2})={\bf
E}\{\rho_{n,p}(\lambda_{1})\rho_{n,p}(\lambda_{2})\}-{\bf
E}\{\rho_{n,p}(\lambda_{1})\}{\bf E}\{\rho_{n,p}(\lambda_{2})\}.
\end{equation}
It is easy to see that the Stieltjes transform of
$\digamma_{n,p}(\lambda_{1},\lambda_{2})$ is
\begin{equation*}
C_{n,p}(z_{1},z_{2})=\iint
\frac{\digamma_{n,p}(\lambda_{1},\lambda_{2})}{(\lambda_{1}-z_{1})
(\lambda_{2}-z_{2})}d\lambda_{1} d\lambda_{2}, \quad {\hbox{  Im}}
\, z_{i}\neq{0}.
\end{equation*}
Applying formally the inversion formula (\ref{c.6roh}), we obtain the following relation
\begin{equation}\label{c.6digammanouv}
\digamma_{n,p}(\lambda_{1},\lambda_{2})=I_{\lambda_{1}}\circ
I_{\lambda_{2}}\{C_{n,p}(z_{1},z_{2})\}.
\end{equation}
In Theorem 2.2, we have found explicitly the leading term of
$C_{n,p}(z_{1},z_{2})$ in the domain $|{\hbox{ Im}}\,z|\geq{2v}$.
However, since the functions  $S$ (2.12) and $T$ (2.13) can  be
continued up to the real axis with respect to the both variables
$z_{1}$ and $z_{2}$, we can apply to the leading term of (2.11) the
operation $I_{\lambda_{1}}I_{\lambda_{2}}$,
$\lambda_{1}\neq{\lambda_{2}}$ to compute formally the ''leading''
term of the density-density correlation function. This means that we
perform first the limit $n,p\rightarrow\infty$ and then the limits
$\epsilon_{1},\epsilon_{2}\downarrow{0}$. This order of limiting
transitions is inverse with respect to that prescribed by the
definition (\ref{c.6digammanouv}).

Let us denote $w_j = w(z_j), \, j=1,2$ and write the identity
\begin{equation*}
\frac{w_{1}-w_{2}}{z_{1}-z_{2}}=\frac{w_{1}w_{2}}{1-v^{2}w_{1}w_{2}}
\end{equation*}
that is an easy consequence of the equation (2.6).
This identity yields relations
\begin{equation}\label{c.6.3}
\epsilon|w(\lambda+i\epsilon)|^{2}=\hbox{Im\,} w(\lambda+i\epsilon)
(1-v^{2}|w(\lambda+i\epsilon)|^{2})
\end{equation}
and $ |w(\lambda+i\epsilon)|^{2}=v^{-2}$ for $\lambda$ such that
$\hbox{Im\,} w(\lambda+i0)>0$. Combining (\ref{c.6.3}) with (1.3),
we obtain that
\begin{equation}\label{c.6.4}
v^{2}[\hbox{Re\,}w(\lambda+i0)]^{2}=\frac{\lambda^{2}}{4v^{2}} \quad
\hbox{and} \quad
v^{2}[\hbox{Im\,}w(\lambda+i0)]^{2}=1-\frac{\lambda^{2}}{4v^{2}}.
\end{equation}
Let us consider the terms of the RHS of (2.11) given by (2.12) and
(2.13). Using (\ref{c.6.4}), we obtain that
\begin{equation*}
I_{\lambda_{1}}\circ
I_{\lambda_{2}}\{S(z_{1},z_{2})\}=-\frac{1}{2v^{2}\pi^{2}[(\lambda_{1}-\lambda_{2})]^{2}}\frac{4v^{2}-\lambda_{1}
\lambda_{2}}{(4v^{2}-\lambda^{2}_{1})^{1/2}
(4v^{2}-\lambda^{2}_{2})^{1/2}}
\end{equation*}
and
\begin{equation*}
I_{\lambda_{1}}\circ
I_{\lambda_{2}}\{T(z_{1},z_{2})\}=\frac{1}{4\pi^{2}v^{8}}\frac{(2v^{2}-\lambda^{2}_{1})(2v^{2}-\lambda^{2}_
{2})}{ (4v^{2}-\lambda^{2}_{1})^{1/2}
(4v^{2}-\lambda^{2}_{2})^{1/2}}.
\end{equation*}
Then for $\digamma_{n,p}(\lambda_{1},\lambda_{2})$ (\ref{c.6.2}) we
get the following formal expression
\begin{equation*}
\digamma_{n,p}(\lambda_{1},\lambda_{2})
=-\frac{1}{\pi^{2}[n(\lambda_{1}-\lambda_{2})]^{2}}\frac{4v^{2}-\lambda_{1}
\lambda_{2}}{(4v^{2}-\lambda^{2}_{1})^{1/2}
(4v^{2}-\lambda^{2}_{2})^{1/2}}
\end{equation*}
\begin{equation}\label{c.6digammafinale}
+\frac{1}{np}\frac{V_{4}}{2\pi^{2}v^{8}}\frac{(2v^{2}-\lambda^{2}_{1})(2v^{2}-\lambda^{2}_
{2})}{ (4v^{2}-\lambda^{2}_{1})^{1/2} (4v^{2}-\lambda^{2}_{2})^{1/2}
}-\frac{1}{n^{2}}\frac{3v^{4}}{2\pi^{2}v^{8}}\frac{(2v^{2}-\lambda^{2}_{1})(2v^{2}-\lambda^{2}_
{2})}{ (4v^{2}-\lambda^{2}_{1})^{1/2} (4v^{2}-\lambda^{2}_{2})^{1/2}
}.
\end{equation}
It is easy to see that in the scaling limit
 $$
 \lambda_{1},\lambda_{2}\rightarrow \lambda, \quad
n(\lambda_{2}-\lambda_{1})\rightarrow s, \eqno (6.7)
$$
one gets equality
$$
\lim_{n(\lambda_{2}-\lambda_ {1})\rightarrow
s}\digamma_{n,p}(\lambda_{1},\lambda_{2})=-\frac{1}{\pi^{2}s^{2}}.
\eqno (6.8)
$$
We see that the terms of the order $O(1/np)$ that depend on the
value $V_4$ disappear in the local scale limit (6.7) and the
density-density correlation function gets the universal form (6.8).
This result can be regarded as an evidence of the fact that the
moderate dilution of Wigner random matrices given by the limit
(2.10) does not change the universality class of the local
eigenvalue statistics.

Condition $3/5<\alpha\le{1}$ (2.10) of Theorem 2.2 is related with
the technical restriction of the approach we use. Indeed, we need
this in the bound (4.26) to estimate the term $\Upsilon$ (4.10) that
corresponds to the last term of the cumulant expansion of the order
5 (see (4.5)). Pushing forward this expansion to the orders higher
than 5, one could consider lower values for $\alpha$. However, this
requires much more cumbersome computations than those of the present
paper.

\vskip 0.5cm

\noindent {\bf Acknowledgements.}  O.K. is grateful for the research project   "Grandes Matrices Al\'eatoires" ANR-08-BLAN-0311-01 
 for the financial support.

\end{document}